\documentclass[fleqn,usenatbib]{mnras}

\usepackage[T1]{fontenc}
\usepackage{ae,aecompl}

\usepackage{graphicx}	
\usepackage{amsmath}	
\usepackage{amssymb}	
\usepackage{hyperref}   
\usepackage{colortbl,comment}
\usepackage{xcolor}
\usepackage{multirow}
\usepackage{multicol}
\usepackage[utf8]{inputenc}

\def\arcsec{$^{\prime\prime}$}
\def\arcmin{$^{\prime}$}

\newcommand{\amm}{NH$_3$}
\newcommand{\cc}{cm$^{-3}$}

\title[Core-Magnetic Field Alignment in Molecular Clouds]{Relative Alignment between Dense Molecular Cores and Ambient Magnetic Field: The Synergy of Numerical Models and Observations }

\author[C.-Y. Chen et al.]{
Che-Yu Chen,$^{1}$\thanks{E-mail: cc6pg@virginia.edu}
Erica A. Behrens,$^{2}$\thanks{Co-second authors}
Jasmin E. Washington,$^{1}\dagger$
Laura M. Fissel,$^{3,4}$
\newauthor
Rachel K. Friesen,$^{3,5}$
Zhi-Yun Li,$^{1}$
Jaime E. Pineda,$^{6}$
Adam Ginsburg,$^{7}$
Helen Kirk$,^{8}$
\newauthor
Samantha Scibelli,$^{9}$
Felipe Alves,$^{6}$
Elena Redaelli,$^{6}$
Paola Caselli,$^{6}$
Anna Punanova,$^{10}$
\newauthor
James Di Francesco,$^{8,11}$
Erik Rosolowsky,$^{12}$
Stella S. R. Offner,$^{13}$
Peter G. Martin,$^{14}$
\newauthor
Ana Chac{\'o}n-Tanarro,$^{15}$
Hope H.-H. Chen,$^{13}$
Michael C.-Y. Chen,$^{12}$
Jared Keown,$^{12}$
\newauthor
Youngmin Seo,$^{16}$
Yancy Shirley,$^{9}$
Hector G. Arce,$^{17}$
Alyssa A. Goodman,$^{18}$
\newauthor
Christopher D. Matzner,$^{5}$
Philip C. Myers$^{18}$
and Ayushi Singh$^{5}$
\\
$^{1}$Department of Astronomy, University of Virginia,
  Charlottesville, VA 22904, USA\\
$^{2}$Department of Astronomy, Ohio State University, Columbus, OH 43210, USA\\
$^{3}$National Radio Astronomy Observatory, Charlottesville, VA 22904, USA\\
$^{4}$Department of Physics, Engineering Physics and Astronomy, Queen’s University, Kingston, ON K7L 3N6, Canada \\
$^{5}$Department of Astronomy \& Astrophysics, University of Toronto, Toronto, ON M5S 3H4, Canada \\
$^{6}$ Max-Planck-Institut für extraterrestrische Physik, Giessenbachstrasse 1, 85748 Garching, Germany\\
$^{7}$ Department of Astronomy, University of Florida, Gainesville, FL 32611, USA \\
$^{8}$ Herzberg Astronomy and Astrophysics, National Research Council of Canada, Victoria, BC, V9E 2E7, Canada \\
$^{9}$ Department of Astronomy, University of Arizona, Tucson, AZ 85721, USA \\
$^{10}$ Ural Federal University, 620002 Mira st. 19, Yekaterinburg, Russia\\
$^{11}$ Department of Physics and Astronomy, University of Victoria, Victoria, BC V8P 5C2, Canada  \\
$^{12}$ Department of Physics, University of Alberta, Edmonton, AB T6G 2E1, Canada \\
$^{13}$ Department of Astronomy, University of Texas at Austin, Austin, TX 78712, USA \\
$^{14}$ Canadian Institute for Theoretical Astrophysics, University of Toronto, Toronto, ON M5S 3H8, Canada\\
$^{15}$ Observatorio Astron{\'o}mico Nacional (OAN-IGN), Alfonso XII 3, 28014, Madrid, Spain \\ 
$^{16}$ Jet Propulsion Laboratory, California Institute of Technology, Pasadena, CA 91109, USA \\
$^{17}$ Department of Astronomy, Yale University, New Haven, CT 06520, USA \\
$^{18}$ Harvard-Smithsonian Center for Astrophysics, Cambridge, MA 02138, USA\\
}

\date{}

\pubyear{2020}
\begin{document}
\label{firstpage}
\pagerange{\pageref{firstpage}--\pageref{lastpage}}
\newgeometry{top=0.7in, bottom=0.6in,left=0.7in, right=0.7in}

\maketitle

\begin{abstract}
The role played by magnetic field during star formation is an important topic in astrophysics.
We investigate the correlation between the orientation of star-forming cores (as defined by the core major axes) and ambient magnetic field directions in 1) a 3D MHD simulation, 2) synthetic observations generated from the simulation at different viewing angles, and 3) observations of nearby molecular clouds. 
We find that the results on 
relative alignment between cores and background magnetic field
in synthetic observations slightly disagree with those measured in fully 3D simulation data, which is partly because cores identified in projected 2D maps tend to coexist within filamentary structures, while 3D cores are generally more rounded.
In addition, we examine the progression of magnetic field from pc- to core-scale in the simulation, which is consistent with the anisotropic core formation model that gas preferably flow along the magnetic field toward dense cores.
When comparing the observed cores identified from the GBT Ammonia Survey (GAS) and {\it Planck} polarization-inferred magnetic field orientations, we find that the relative core$-$field alignment has a regional dependence among different clouds. More specifically, we find that dense cores in the Taurus molecular cloud tend to align perpendicular to the background magnetic field, while those in Perseus and Ophiuchus tend to have random (Perseus) or slightly parallel (Ophiuchus) orientations with respect to the field. We argue that this feature of relative core$-$field orientation could be used to probe the relative significance of the magnetic field within the cloud.
\end{abstract}

\begin{keywords}
ISM: magnetic fields -- ISM: structure -- MHD -- polarization -- stars: formation
\vspace{-.7in}
\end{keywords}
\restoregeometry


\section{Introduction}

Stars form within molecular clouds (MCs) through the gravitational collapse of dense cores \citep{Shu1987}.  Prestellar core formation in MCs is therefore an important issue in theoretical studies of star formation, because these cores are the immediate precursors of protostars \citep{Andre2009}.
It is generally accepted that magnetic effects, in combination with turbulence and gravity, are one of the key agents affecting the dynamics of the star-forming process at all physical scales and throughout different evolutionary stages \citep{BP2007PPV, MO2007}.
In particular, while the cloud-scale magnetic fields could limit compression in turbulence-induced shocks and regulate gas flows toward overdense regions \citep{1956MNRAS.116..503M, 1985prpl.conf..320M, Heyer+08,2015ApJ...810..126C, 2017ApJ...847..140C}, 
the interconnected core-scale magnetic field is expected to be important in regulating the gas dynamics within cores via e.g., removing angular momentum in collapsing cores \citep[see review in][]{Li2014PPVI}. 

Observationally, the polarized thermal emission from dust grains at infrared/sub-mm wavelengths has been considered as a reliable tracer of plane-of-sky magnetic field morphology in dense star-forming clouds \citep[e.g.,][]{1984ApJ...284L..51H, Heiles1993PP, Novak1997}, because the elongated grains are generally recognized to have a preferential orientation with their longer axes perpendicular to the local magnetic field \citep{DG1951, Lazarian07, LH07}. With advanced polarimetric technologies and instruments, linear polarization observations have recently become a powerful tool to investigate the magnetic effects during star formation at various physical scales \citep[e.g.,][]{Matthews09,Hull+14,Cox15,2016ApJ...824..134F,Haifeng16}. 
Specifically, the all-sky map of polarized emission from dust at sub-mm wavelengths produced by the {\it Planck} satellite revealed Galactic magnetic field structures, which provide new insight into understanding the magnetic effects in nearby star-forming MCs \citep{planckXIX,pXXXIII,pxxxv16}.

Dense molecular cores are observed in dust continuum and molecular lines. Recent results suggest that the cores are associated with filamentary structures within star-forming MCs \citep{adw+14}, which agrees with various simulations of turbulent clouds \citep[e.g.,][]{2014ApJ...789...37V,2014ApJ...791..124G,2014ApJ...785...69C,2015ApJ...810..126C,GO15}. 
Millimeter and sub-mm continuum surveys of nearby clouds showed that prestellar cores have masses $\sim 0.1-10$~M$_\odot$ and sizes $\sim 0.01-1$~pc \citep[e.g.,][]{Enoch+06,Johnstone06,Konyves+10,Kirk+13}, while the gas dynamics at core scales have been investigated in spectral line observations using dense gas tracers like N$_2$H$^+$, H$^{13}$CO$^{+}$, or C$^{18}$O \citep[e.g.,][]{2009ApJ...691.1560I,2009ApJ...699..742R,Pineda+15,Punanova+18,GBTcore19}.
In particular, analyses of core kinematics from the recent Green Bank Ammonia Survey \citep[GAS;][]{fp+17} shows that many dense cores are not bound by self-gravity, but are pressure-confined \citep{Kirk+17,2017ApJ...850....3K,HopeGAS19,kerr_2019}.

Though the interaction between dense gas and magnetic fields during the formation and evolution of prestellar cores has been extensively studied in various theoretical models \citep[e.g.,][]{Ostriker+99,Basu+04,LiNakamura04,McKee+10,KudohBasu2011,2014ApJ...785...69C,2015ApJ...810..126C}, systematic observational investigations were relatively lacking because of a dearth of magnetic field observations before the advent of several new instruments like BLASTPol \citep{2012SPIE.8444E..15P}, {\it Planck}, and ALMA.
Recently, \cite{2014ApJ...791...43P} looked at the orientation of cores in Lupus I identified within {\it Herschel} SPIRE maps \citep{2013A&A...549L...1R} compared to the orientation of the Lupus I filament and the average magnetic field derived from BLASTPol observations \citep{2014ApJ...784..116M}. They found no correlation between the core angle and the magnetic field, but they were unable to compare the core orientation to the direction of the ambient magnetic field near the cores.

Following recent achievements by several research groups on statistically characterizing the significance of magnetic effects in star-forming regions using observable information from both polarimetric and spectral measurements \citep[e.g.,][]{Soler+13,2015ApJ...807....5F,2016ApJ...829...84C,2019MNRAS.485.3499C,Fissel+19}, here we present our investigations on the relative orientations between dense cores and magnetic fields in both simulations and observations.
The goal of this study is to determine whether or not the surrounding magnetic fields have significant impacts on the structures of star-forming cores. 

The outline of the paper is as follows. We describe our numerical and observational approaches in Section~\ref{sec::methods}, where we also discuss the methods we adopt to identify cores (Section~\ref{sec::coreId}). 
The main results are included in Section~\ref{sec::results}.
We present our results from simulated cores in Section~\ref{sec::simcores}. Synthetic observations are investigated in Section~\ref{sec::synobs}, and measurements from real observation data are shown in Section~\ref{sec::obscores}. Further discussions are presented in Section~\ref{sec::disc}, where
we cross-compare 3D vs.~2D core identifications from the original simulation and the synthetic observations (Section~\ref{sec::aspRatio}).
We also investigate and compare the results from synthetic and real observations, and link them to the general properties of the observed clouds (Section~\ref{sec::cloudB}). Our conclusions are summarized in Section~\ref{sec::sum}.

\section{Methods}
\label{sec::methods}

\subsection{Simulation and Synthetic Observations}
\label{sec::sims}

\begin{figure*}
\begin{center}
\includegraphics[width=0.35\textwidth]{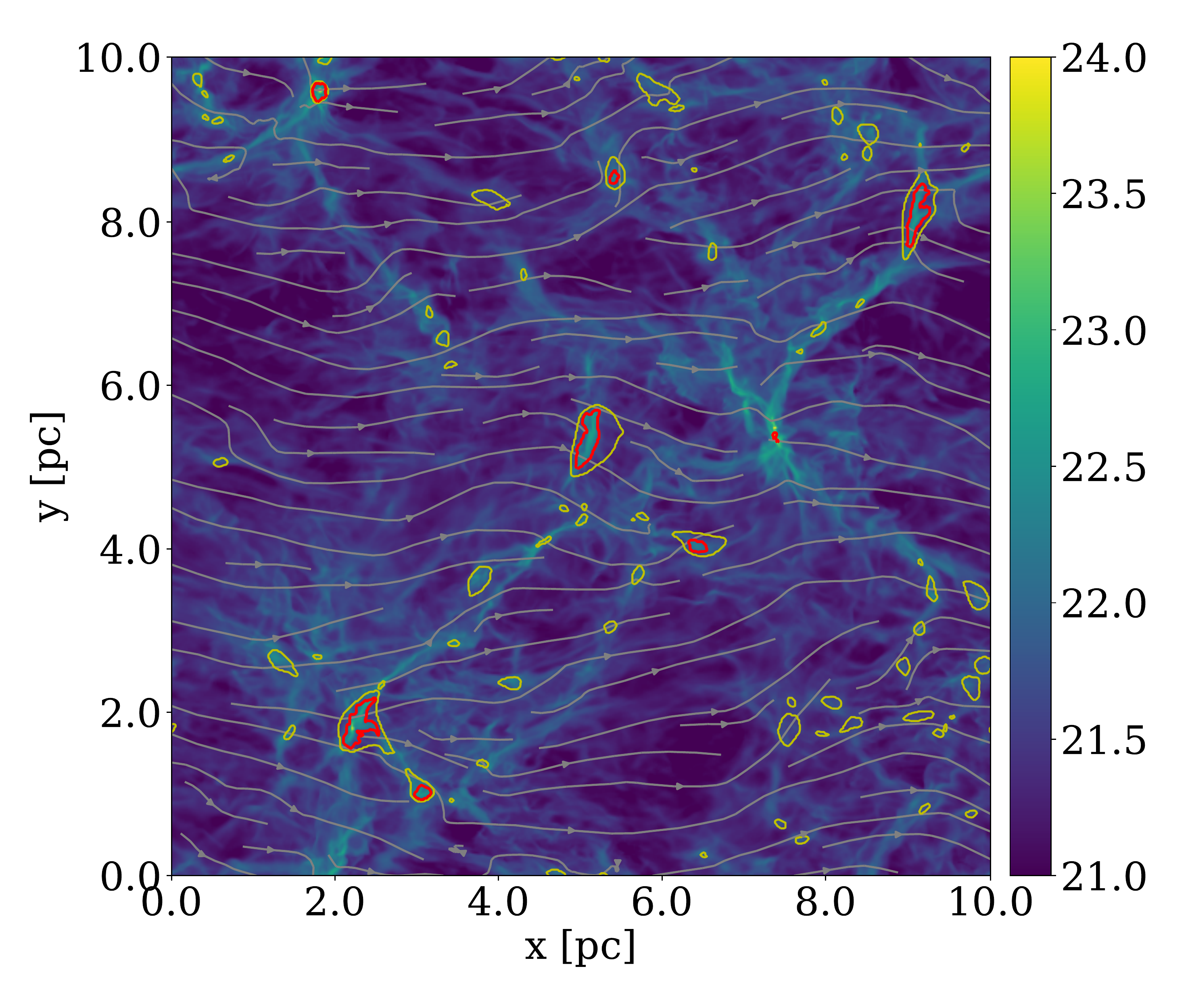}
\includegraphics[width=0.35\textwidth]{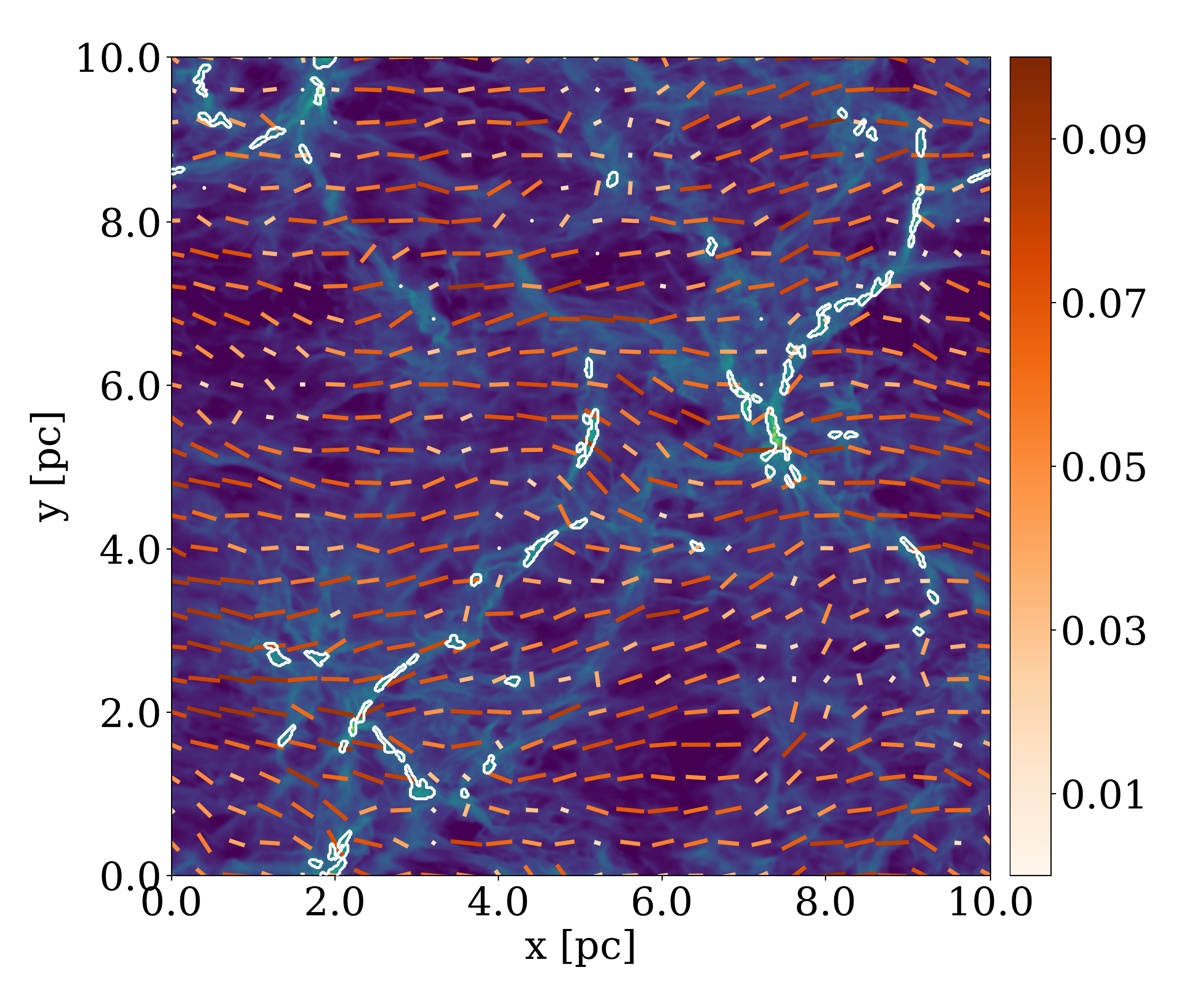}
\includegraphics[width=0.276\textwidth]{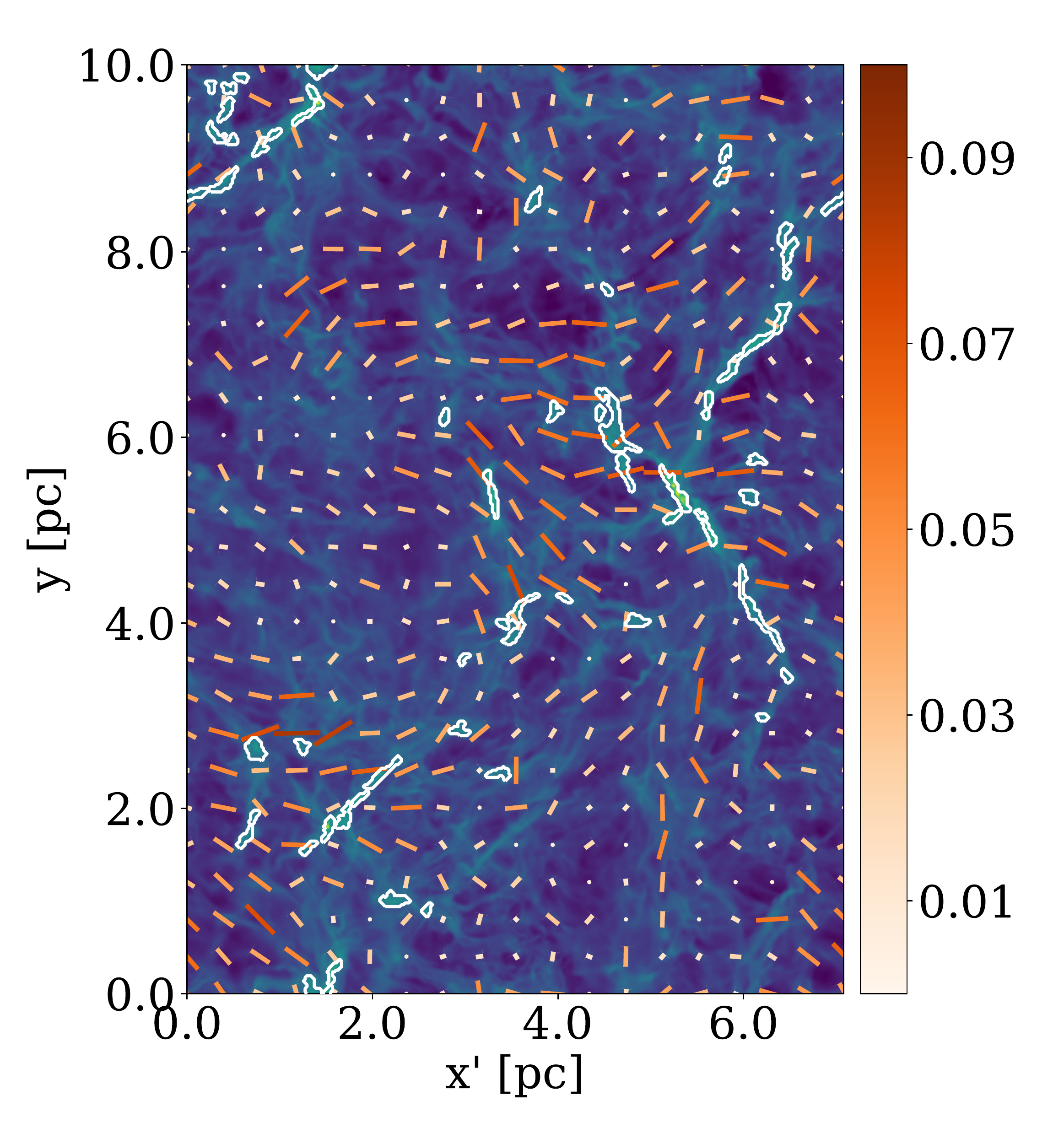}
\caption{Identified cores ({\it contours}; see Section~\ref{sec::coreId} for how cores are defined) overlaid on integrated column density of atomic hydrogen (in the same $\log$ scale for all three panels; see the colorbar of the left panel). 
{\it Left:} Cores identified in 3D, both bound ({\it red contours}) and unbound ({\it yellow contours}). Grey streamlines represent density-weighted magnetic field structure. {\it Middle and right panels:} Cores ({\it white contours}) identified in projected column density maps with synthetic polarization direction (along the inferred magnetic field; {\it orange segments}), for both face-on view ({\it middle}) and $45^\circ$ viewing angle ({\it right}). The color bars in the center and right panels indicate the polarization fraction. }
\label{fig::simcores}
\end{center}
\end{figure*}

The simulation considered in this study was first reported in \citet[their model A]{2018MNRAS.474.5122K} and \citet[their model L10]{2019MNRAS.485.3499C} and is summarized here. This 10\,pc-scale simulation, considering a magnetized shocked layer produced by plane-parallel converging flows, was particularly designed to follow the formation of dense structures in cloud-cloud collision.
Previous studies have suggested that this mechanism efficiently produces star-forming regions \citep[see e.g.,][]{2000ApJ...532..980K,2006ApJ...643..245V,2008ApJ...674..316H,2009MNRAS.398.1082B,2013ApJ...774L..31I}. The average amplitude of the velocity perturbation within the convergent flows was chosen by setting the virial number of the colliding clouds equal to 2. The resulting post-shock layer (approximately $\sim 2-3$~pc thick), wherein the dense cores considered in this study form, is strongly magnetized with plasma beta $\beta \equiv 8\pi\rho{c_s}^2/B^2 \approx 0.1$ and moderately trans- to super-Alfv\'enic with the median value of Alfv\'en Mach number\footnote{Note that this value slightly differs from that quoted in \cite{2018MNRAS.474.5122K}, which is due to different methods of defining the post-shock region adopted in their study and this one.} $\langle {\cal M}_{\rm A}\rangle \equiv \langle v / v_{\rm A}\rangle \sim 1.5$, where $v$ is the gas velocity and $v_{\rm A} \equiv B/\sqrt{4\pi\rho}$ is the Alfv\'en speed in the cloud.
This simulation was chosen for this study because it has a resolution comparable to the observation data considered here ($\sim 0.02$~pc; see Section~\ref{sec::GAS}), and its physical properties (volume/column densities, gas velocities, etc.) are very close to those measured in typical star-forming regions \citep[e.g.,][]{CLASSy2014}. In fact, it has been shown that the synthetic observations generated from this simulation have similar polarimetric features to those found in real observations \citep{2019MNRAS.490.2760K}. 
The projected column density and the density-weighted average magnetic field of the simulation are shown in the left panel of Figure~\ref{fig::simcores}. Note that the magnetic field is roughly aligned along the $x$-axis.

We generated the synthetic observations (which will be investigated in Section~\ref{sec::synobs}) by projecting the simulation box along either the $z$-direction (face-on case) or the line of sight at $45^\circ$ from the $z$-axis toward the $x$-axis ($45^\circ$ case). The column density can be calculated directly by integrating the number density along the line of sight, and we adopted 10\% abundance of helium to convert the simulated neutral molecular density to atomic hydrogen density, i.e., $n_{\rm neutral} \approx n_{\rm H_2} + n_{\rm He} = 0.6 n_{\rm H}$.
There is no noise added.

To derive the synthetic polarimetric measurement, we followed previous work \citep[e.g.,][]{1985ApJ...290..211L,2000ApJ...544..830F,2018MNRAS.474.5122K} 
and assumed viewing from $+z$ direction:
\begin{equation}
q = \int n \frac{B_y^2 - B_x^2}{B^2} ~ds,\ \ u = \int n \frac{2B_xB_y}{B^2} ~ds,
\label{poldef2}
\end{equation}
where $n$, $B_x$, $B_y$, $B_z$, and $B$ represent the number density, $x$, $y$, $z$ components and the absolute value of the total magnetic field at each location along the line of sight $ds$, respectively. 
Here, $q$ and $u$ (in units of column density) trace the orientation dependence of polarization arising from the grain alignment with respect to the magnetic field and are proportional to the predicted polarized intensities $Q$ and $U$ by an assumed-constant scale factor $\kappa_\nu B_\nu(T_d) p_0$, where $\kappa_\nu$ is the dust opacity, $B_\nu(T_d)$ is the Planck function at dust temperature $T_d$, and $p_0$ is a polarization factor relating to the grain shape, composition, and orientation \citep{PlanckXX}.\footnote{We adopt $p_0 = 0.1$ in this study, in the range $0.1 - 0.3$ found empirically from radio/sub-mm observations \citep[see e.g.,][]{2016ApJ...824..134F,PlanckXX}.  The actual value might vary within the cloud, as discussed by, e.g.,~\cite{2019MNRAS.490.2760K}.}
Note also that because we are not trying to resolve the polarization directions within the dense cores, just the ambient materials, we did not consider the depolarization effect in the high-density regime, which becomes significant at $n \gtrsim 10^6$~cm$^{-3}$ \citep[see e.g.,][]{2012A&A...543A..16P, 2019MNRAS.490.2760K}. 
The polarization fraction is therefore given by
\begin{equation}
p = p_0 \frac{\sqrt{q^2 + u^2}}{N - p_0 N_2},
\end{equation}
where 
$N$ is the column density, and $N_2$ is a correction to the column density which depends on the inclination of the dust grains:
\begin{equation}
N_2 = \int n\left(\cos^2 \gamma - \frac{2}{3}\right) ~ds.
\end{equation}
Here, $\gamma$ is the inclination angle of the magnetic field relative to the plane of sky, i.e.,~$\cos^2\gamma = (B_x^2+B_y^2)/B^2$.
The inferred polarization angle on the plane of sky is therefore
\begin{equation}
\chi = \frac{1}{2} \text{arctan2}(u,q),
\label{eq:synchi}
\end{equation}
which is measured clockwise from North. 
Note that while in general the observed polarization orientation is $90^\circ$ from the inferred magnetic field, the synthetic polarization as derived from the above equations is along the inferred magnetic field. 

The two synthetic observations are illustrated in Figure~\ref{fig::simcores} (middle and right panels), which shows the projected atomic hydrogen column density maps along with line segments indicating the inferred magnetic field orientations from polarization.

\subsection{Actual Observations}
\subsubsection{GAS data}
\label{sec::GAS}

Dense molecular cores were identified in integrated intensity maps of ammonia (\amm) (1,1) inversion transition emission as observed by the Green Bank Ammonia Survey (GAS). We refer the reader to the first GAS data release in \citet{fp+17} for a full description of the survey, its goals and methods of data aquisition, reduction, and analysis methods, but here describe briefly the key points. 

\amm\, is an excellent tracer of moderate-to-high density gas, as it is formed and excited at densities $n \gtrsim 10^4$~\cc. Observations of the \amm\ (1,1) through (3,3) inversion line transitions at $23.7~\mathrm{GHz} - 23.9~\mathrm{GHz}$ were performed from January 2015 through March 2016 at the Robert C. Byrd Green Bank Telescope (GBT) using the 7-pixel K-band Focal Plane Array (KFPA) and the VErsatile GBT Astronomical Spectrometer (VEGAS). All observations were performed using the GBT's On-The-Fly mode, where the telescope scans in Right Ascension or declination, and the spacing of subsequent scans is calculated to provide Nyquist spacing and consistent coverage of the desired map area. Individual maps were usually 10\arcmin\, $\times$ 10\arcmin\, in size, and multiple such maps were combined to provide full coverage across the observed regions. As described more fully in \citet{fp+17}, the \amm\ (1,1) integrated intensity maps were created by summing over spectral channels with line emission, as determined through pixel-by-pixel fitting of the \amm\ hyperfine structure. 

In this paper, we examine eleven distinct regions within three clouds: B1, L1448, L1451, L1455, NGC\,1333, and IC348 in the Perseus molecular cloud; L1688 and L1689 in the Ophiuchus molecular cloud; and B18, HC2, and B211/213 in the Taurus molecular cloud. B211/213 was not included in the GAS project, as it was observed previously at the GBT with similar sensitivity and spectral setup \citep{seo_2015}. All three clouds are relatively nearby ($d \sim 135$~pc to $\sim 300$~pc), such that the $\sim 31$\arcsec\ GBT beam at 23.7~GHz subtends $\sim 0.02~\mathrm{pc} - 0.04$~pc, sufficient to resolve dense cores that are typically $\sim 0.1$~pc in size. 

\subsubsection{Planck data}
\label{sec::planck}

We used maps of linearly polarized dust emission at 353\,GHz from the {\em Planck} Satellite to infer the pc-scale magnetic field orientation over large scales towards the three nearby clouds where we have \amm\ data from the GAS survey: Taurus, Perseus and Ophiuchus. 
To enhance the signal-to-noise ratio, the {\it Planck} Stokes $I$, $Q$, and $U$ maps were smoothed to a resolution of 15\arcmin{} FWHM following the covariance smoothing procedures described in \cite{planckXIX}, as has been done in \cite{pxxxv16} and \cite{Soler19}.
The 15\arcmin\ FWHM corresponds to a linear resolution of $\sim 0.6$\,pc for Taurus and Ophiuchus ($d\sim 140$~pc), and $\sim 1.2$\,pc for Perseus ($d\sim 300$~pc).

Similar to the selection criteria applied in \citet{pxxxv16} we only include measurements where the {\em Planck}-derived~column density $N_{\mathrm H}$~is above the RMS column density in a reference region of diffuse ISM at the same Galactic latitude.  We also require that the linearly polarized intensity $P$~obeys
\begin{equation}
    P > 2\,P_{\rm ref},
\end{equation}
where $P$ is the quadrature sum of Stokes $Q$~and $U$
\begin{equation}\label{eqn:polmask}
    P = \sqrt{Q^2\,+\,U^2},
\end{equation}
and $P_\mathrm{ref}$~is the corresponding RMS value of $P$ in the reference region.  Finally, we require at least a 3\,$\sigma$~detection of $P$. 

To infer the magnetic field orientation projected on the sky, $\hat{B_{\mathrm POS}}$, we measure the orientation of linear sub-mm polarization and rotate it by 90$^{\circ}$:
\begin{equation}\label{eqn:bangle}
\hat{B_{\mathrm POS}}\,=\,\frac{1}{2}\,\arctan^{-1}\left(-U,Q\right) + \frac{\pi}{2}.
\end{equation}
Note that the difference between Equations~(\ref{eq:synchi}) and (\ref{eqn:bangle}) comes from the need to convert the {\it Planck} $Q$ and $U$ maps, which use the HEALPix standards\footnote{A position angle of zero is oriented towards Galactic north, and the position angle increases as the orientation rotates counter clockwise or East of North on the sky.} for coordinate systems \citep{planckXIX}.

\subsection{Core Identification}
\label{sec::coreId}

In this section we describe the algorithms we chose to identify dense cores from 3D space (simulation) and 2D projected maps (synthetic and real observations). Since the main purpose of this study is to determine the orientation of dense core (which is defined by its major axis) and how it correlates with magnetic field structure, we exclude rounded cores with ratios of minor to major axes ($b/a$ for 3D cores where $a$ represents the major axis and $b$ the second-longest axis) larger than 0.8 to 
remove cores that have poorly defined or uncertain major and minor axes.

\begin{table*}
\begin{center}
  \caption{Summary of the cores considered in this study. $\langle\rangle$ represents the median value of a property, and $\measuredangle [u,v]$ denotes the angle between parameters $u$ and $v$. Note that angles in simulation data are measured in 3D, while angles in synthetic and GAS observations are measured in 2D. For all cores, we use $a$, $b$, and $c$ to represent the lengths of the three axes of a core. For 3D cores, $a$ corresponds to the major axis, and $c$ is the minor axis. For cores in synthetic observations and in GAS data, there are only two axes, and thus $b$ represents the minor axis. The noise level $\sigma$ is measured in K~km\,s$^{-1}$ for GAS observations. }
	\label{tab:cores}
\begin{tabular}{l | cccccccc } 
		\hline
		\multirow{2}{*}{case} & noise & \# Cores & $\langle R_{\rm core}\rangle$ & \multirow{2}{*}{$\langle b/a\rangle$} & \multirow{2}{*}{$\langle c/a\rangle$} & $\langle\measuredangle [B_{\rm bg}, a]\rangle$ & K-S test on & K-S test,\\
		& level $\sigma$ & Identified & [pc] & & & [deg] & $\measuredangle [B_{\rm bg}, a]$ & $p$-value\\
		\hline
		\multicolumn{9}{c}{\bf simulation} \\
		\hline
		unbound & $-$ & 78 & 0.048 & 0.53 & 0.29 & 52.1 & 0.15 & 0.25 \\
		bound & $-$ & 8 & 0.072 & 0.45 & 0.22 & 62.1 & 0.21 & 0.86\\
		\hline
		\multicolumn{9}{c}{\bf synthetic observation} \\ 
		\hline
		face-on & $9.5\times 10^{20}$ cm$^{-2}$& 82 & 0.060 & 0.35 & $-$ & 52.7 & 0.17 & 0.15\\
		$45^\circ$ & $1.7\times 10^{21}$ cm$^{-2}$ & 68 & 0.078 & 0.39 & $-$ & 49.8 & 0.15 & 0.33\\
		\hline
		\multicolumn{9}{c}{\bf GAS observation} \\ 
		\hline
		{\bf Ophiuchus} & $-$ & {\bf 38} & {\bf 0.035}& {\bf 0.53} & $-$ & {\bf 36.3} & {\bf 0.15} & {\bf 0.54}\\
		-- L1688 & 0.093 & 34 & 0.037 & 0.54 & $-$ & 36.3 & $0.15$ & 0.55\\
		-- L1689 & 0.101& 4 & 0.032 & 0.75 & $-$ & 45.3 & $0.15$ & 1.00\\
		\hline
		{\bf Perseus} & $-$ & {\bf 80} & {\bf 0.057}& {\bf 0.54} & $-$ & {\bf 54.7} & {\bf 0.14} & {\bf 0.28}\\
		-- B1 & 0.072 & 20 & 0.051 & 0.62 & $-$ & 54.7 & $0.21$ & 0.41\\
		-- L1455 & 0.069 & 12 & 0.069 & 0.55 & $-$ & 29.0 & $0.42$ & 0.03\\
		-- NGC\,1333 & 0.072 & 28 & 0.052 & 0.56 &  $-$ & 56.3 & $0.19$ & 0.39\\
		-- IC348 & 0.070& 20 & 0.060 & 0.52 & $-$ & 63.7 & $0.34$ & 0.03\\
		\hline
		{\bf Taurus} & $-$ & {\bf 35} & {\bf 0.044}& {\bf 0.50} & $-$ & {\bf 63.7} & {\bf 0.26} & {\bf 0.04}\\
		-- B18 & 0.075& 5 & 0.047 & 0.70 & $-$ & 43.7 & $0.40$ & 0.34\\
		-- HC2 & 0.101& 12 & 0.030 & 0.47 & $-$ & 75.6 & $0.59$ & 0.00\\
		-- B211/213 & 0.151 & 18 & 0.049 & 0.51 & $-$ & 54.0 & $0.22$ & 0.42\\
		\hline
	\end{tabular}
	\end{center}
\end{table*} 

\subsubsection{Cores in 3D space}
\label{sec::simCoresId}

We use the {\it GRID} core-finding algorithm to identify simulated cores in 3D space, which uses the largest closed gravitational potential contours around single local minima as core boundaries, and applies principal component analysis (PCA) to define the three axes of each core, denoted as $a$, $b$, and $c$ from the longest to the shortest. The original {\it GRID} core-finding routine was developed by \cite{2011ApJ...729..120G}, while the extensions to measure the magnetic and rotational properties were implemented by \cite{2014ApJ...785...69C,2015ApJ...810..126C,2018ApJ...865...34C}. 
Note that the gravitational potential morphology is usually more smooth and spherically symmetric than the actual density structure, which makes the identified cores more rounded compared to the density isosurfaces. Nevertheless, though these cores defined by gravitational potential contours may not completely reflect the real-time gas distribution, they represent the effective boundaries of dense cores. 
The identified cores are illustrated in Figure~\ref{fig::simcores} (left panel).
Cores with less than 8 voxels total are excluded to ensure better measurement of the major and minor axes.
We also note that the average densities inside these cores are $n_{\rm H}\sim 10^5$~cm$^{-3}$, comparable to those traced by \amm{} \citep[typical excitation density $\sim 10^3$~cm$^{-3}$; see e.g.,][]{2015PASP..127..299S}.

We further consider the energy balance within each core by measuring the thermal, gravitational, and magnetic energies inside the identified core boundaries. 
For the sub-regions within cores that satisfy $E_{\rm th} + E_{\rm grav} + E_B < 0$, these sub-cores are characterized as gravitationally bound cores. These cores are shown as red contours in Figure~\ref{fig::simcores} (left panel).
Note that for all cores classified as unbound at their largest closed gravitational potential contours (yellow contours in Figure~\ref{fig::simcores}, left panel), there are only a few cores with bound sub-cores somewhere within (8 over 78; see Table~\ref{tab:cores}). 
In our analysis (discussed in Section~\ref{sec::simcores}), we consider both bound and unbound cores, but use different symbols to remind the readers that these two types of cores should be considered separately.

For each core, we measure the average magnetic field direction within the identified core boundary and refer to it as the core-scale magnetic field, $B_{\rm core}$. 
However, the typical core size is $\lesssim 0.1$~pc, much smaller than the resolution of {\it Planck} polarization maps.
The polarization-inferred magnetic field traced by {\it Planck} is therefore not the core magnetic field; instead, {\it Planck} traces the large-scale magnetic field at the position of the core.
We thus define the {\it background} magnetic field for each core, $B_{\rm bg}$, as the average field within the $0.5$~pc-wide cube centered at the core's density peak but excluding the core region.\footnote{Indeed, the 0.5~pc-scale is still within a single {\it Planck} resolution element, but we have tested that measuring $B_{\rm bg}$ at 1~pc-scale did not make significant difference to our results and conclusions.} 
These two quantities are compared in Section~\ref{sec::simcores}.
Note that these two magnetic fields are both direct averages of the volumetric quantity within the specified region, and thus could differ from the observed, density-weighted fields, especially within overdense structures like the cores. Since we only consider the pc-scale polarimetric observations in this study, the difference between volume- and density-weighted fields is not critical.

\subsubsection{Cores in real and synthetic observations}
 \label{sec::obsCoreID}

We identify cores in both the observed data from GAS and synthetic observations using the {\tt astrodendro}\footnote{\url{http://www.dendrograms.org/}} algorithm, which is a Python package to identify and categorize hierarchical structures in images and data cubes.  

The dendrogram algorithm characterizes hierarchical structure by identifying emission features at successive isocontours in emission maps or cubes (leaves), and tracking the intensity or flux values at which they merge with neighboring structures (branches and trunks). The algorithm thus requires the selection of the emission threshold (so that all identified structures must have emissions above this value) and contour intervals (as a step size when looking for successive contours) used to identify distinct structures, which are usually set to be some multiple of the RMS noise properties of the data. 

For the observational data, each region has slightly different noise properties depending on the conditions in which the maps were observed. Furthermore, due to the sparse spacing of pixels in the KFPA, the outer edge of each map receives less integration time overall. The resulting distribution of RMS noise values per pixel, determined in non-line channels within each data cube, follows a skewed Gaussian. We therefore fit skewed Gaussian curves to the noise distribution of each region to determine the peak noise value as well as the width of the distribution, $\sigma$. 
The values of $\sigma$ for individual regions are listed in Table~\ref{tab:cores}.

\begin{figure}
\begin{center}
    \includegraphics[width=\columnwidth]{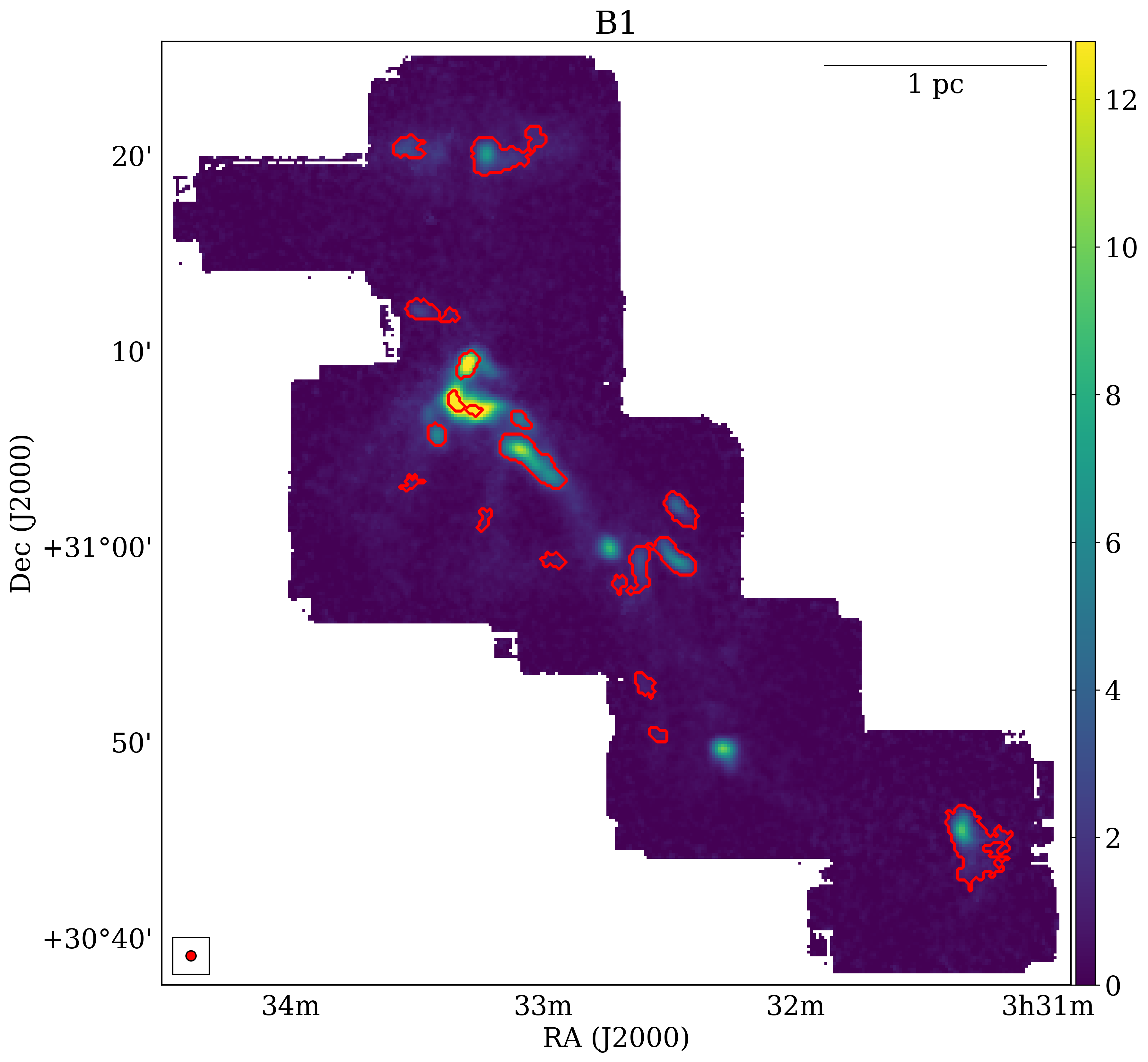}
    \caption{Cores (dendrogram leaves that passed our criteria; {\it red contours}) identified in \amm\ (1,1) integrated intensity emission (K~km\,s$^{-1}$; colormap) toward the B1 region in Perseus. The beam is shown in the lower left corner.} 
    \label{fig:GAScore}
\end{center}
\end{figure}

We create dendrograms of the regions observed while varying the threshold and contour interval in units of the $\sigma$ value for each region. Comparing the identified structures with the integrated intensity and noise maps, we determine that for most regions, a threshold of $9\,\sigma$ and a contour interval of $3\,\sigma$ produced a complete catalog of cores while avoiding identifying noise peaks as real structures. The Ophiuchus regions L1688 and L1689 required a slightly greater contour interval of 5\,$\sigma$, however, due to greater variation in the noise properties across their maps. Indeed, a threshold of 9\,$\sigma$ is greater than what is typically used with \texttt{astrodendro}, but we find it provides a more conservative estimate of real structures given the varying noise properties of the maps. In addition, in this study we are interested only in the dendrogram `leaves', which represent the dense molecular cores that may eventually form (or have already formed) stars. These structures tend to be brighter in \amm\ emission and the higher threshold does not impact negatively our ability to detect them. 

For each core, we obtain measurements of the core position in R.A. and Dec., total area, flux, major and minor axes, and the position angle in degrees counter-clockwise from the +x axis. 
The core major and minor axes are calculated from the intensity-weighted second moment in the direction of the greatest elongation (major), or perpendicular to the major axis (minor). 
These RMS sizes are then scaled up by a factor of $\sqrt{8\ln{2}}$ to get the FWHM of the intensity profile assuming a Gaussian distribution \citep{2006PASP..118..590R}.
Besides the requirement that the ratio between minor and major axes must be smaller than 0.8,
we furthermore apply two additional criteria to the final core catalogs. First, we exclude cores with total area smaller than the size of the GBT beam ($\sim 31$\arcsec\ at 23.7~GHz). Second, we remove cores with total flux values that are less than a factor of 1.25 times the expected flux for a structure with a flat flux profile at the threshold value; i.e., we require an emission peak within the core.
Figure \ref{fig:GAScore} illustrates the resulting cores identified in the B1 region in Perseus observed with GAS; maps of other regions are included in Appendix~\ref{sec::GASmaps}.

For comparison with the {\it Planck} data, we convert the core properties to galactic coordinates. Due to the drastic difference in resolution between the two data sets, each core consists of only a few pixels in the {\it Planck} maps. 
In addition, no cores in the sub-regions L1448 and L1451  in Perseus passed the noise cuts we made to the {\it Planck} data (see Section~\ref{sec::planck}). 
These two regions are therefore excluded from our analysis. Table~\ref{tab:cores} summarizes the final core count and the median values of the core size and minor/major axis ratio in each region.

Note that some cores identified in the \amm{} data are protostellar in nature, i.e.,~some of the \amm{} cores are associated with protostars. This is different from the dense cores identified from the simulation, which are all starless or prestellar because the sink particle routine was not included in the simulation. However, we would like to argue that only the innermost parts of the protostellar cores are directly participating in protostellar evolution, and thus the gas structure at envelope/core scale is not significantly affected at this early protostellar stage \citep[see e.g.,][]{Foster+09,Pineda+11,Pineda+15}. 
Since these \amm{} cores are not small ($\sim 0.03-0.05$~pc; see Table~\ref{tab:cores}), this discrepancy in evolutionary stages of cores is not critical in our analysis on the core geometry. However, we also note that the simulation considered in this study does not include outflows, which could in principle affect the morphology of the core. This may explain why our results from synthetic observations do not agree with regions with active star formation and significant feedback (Section~\ref{sec::obscores}); see Section~\ref{sec::cloudB} for more discussions on regional differences among clouds.

For the synthetic observations, 
we apply \texttt{astrodendro} on the projected column densities of atomic hydrogen to identify cores, which are displayed in Figure \ref{fig::simcores}. 
We averaged over areas of the column density maps with little structure to extract a synthetic noise level $\sigma$ for each synthetic observation, which are also listed in Table~\ref{tab:cores}. 
Though we did not include any noise when generating the synthetic column density maps and thus the $\sigma$ values measured here are not truly comparable to the observational noise, we note that the main purpose here is to define a reasonable background threshold for \texttt{astrodendro} to identify structures above this value.
We therefore set the threshold to be $1.4\times 10^{22}$~cm$^{-2}$ (roughly $\sim 17\,\sigma$ and $\sim 8\,\sigma$ for face-on and $45^\circ$ cases, respectively) for both synthetic maps.
The identified core boundaries, however, should not depend on the contour interval as long as the interval is small enough. We thus set the contour interval for \texttt{astrodendro} to be $3\,\sigma$, similar to what we adopted for observational data. In addition, we consider only \texttt{astrodendro} leaves with number of pixels $>10$ to ensure an accurate core orientation measurement.

\section{Results}
\label{sec::results}

In this section, we present our comparisons between core orientations and multi-scale magnetic fields (core-scale measured locally vs.~larger, pc-scale measured from the background) in the simulation (Section~\ref{sec::simcores}), in synthetic observations of the simulation in projection (Section~\ref{sec::synobs}), and in actual observations (Section~\ref{sec::obscores}). Note that we study the statistics using the cosine values for angles defined in 3D space, while 2D angles are compared in degrees. This is because two random vectors in 3D are more likely perpendicular than aligned, and thus the distribution function of random 3D angles is flat in cosine, not degree. 
We provide a simple numerical justification of such choices in Appendix~\ref{sec::apx}. 
Also note that the relative angle between the core orientation and magnetic field (or polarization) direction is limited to be within $[0,90]$ degree, or $\cos({\rm angle}) \in [0,1]$ to account for the degeneracy of angles larger than $90^\circ$.

\subsection{Core--Magnetic Field Alignment in Simulations}
\label{sec::simcores}

\subsubsection{Magnetic fields within dense cores and at pc-scales: the local$-$background magnetic field alignment}
\label{sec::simBcBbg}

As described in Section~\ref{sec::coreId}, we measure the average magnetic field direction at two scales for each core, locally at the core-scale ($B_{\rm core}$) and at pc-scale (the {\it background}, $B_{\rm bg}$). Whether or not the direction changes significantly from one to another could provide information on the progression of the magnetic field during the process of dense core formation.

\begin{figure*}
    \centering
    \includegraphics[width=0.8\textwidth]{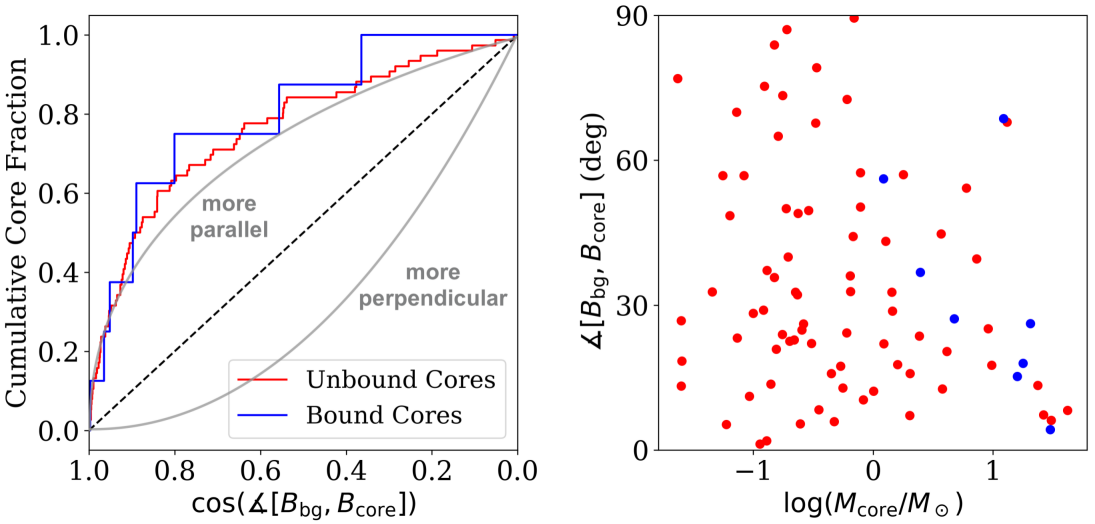}
    \caption{{\it Left:} The cumulative distribution of relative angles between the core-scale and background magnetic fields for the simulated cores identified in 3D, $\measuredangle[B_{\rm bg}, B_{\rm core}]$, measured in cosine. The black dashed diagonal line represents random distribution. Note that the $x$-axis is inverse (from 1 to 0) so that it corresponds to angles from $0^\circ$ to $90^\circ$. {\it Right:} The relative angle (in degree) as a function of core mass, for both unbound ({\it red}) and bound ({\it blue}) cores. }
    \label{fig:Bbg}
\end{figure*}

Figure~\ref{fig:Bbg} (left panel) shows the cumulative distribution function (as a step function) of the relative angle between the local and background magnetic field (in $\cos({\rm 3D\ angles})$) for all cores identified in the simulation.
As shown in the sharp rise at the smaller angles (cosine $\sim 1$) in the step function, $B_{\rm core}$ and $B_{\rm bg}$ tend to align parallel to one another, although a few bound and unbound cores have almost perpendicular alignment. 
Comparing with a random distribution (the straight diagonal line in the plot), the Kolmogorov-Smirnov (K-S) statistic of the local--background magnetic field alignment for the unbound and bound cores are 0.45 ($p$-value $\approx 0.0$) and 0.55 ($p$-value $\approx 0.01$), respectively, indicating that they are far from randomly distributed.
This suggests that the core-scale magnetic field structure is not significantly altered during the mass-gathering process of the dense cores in this simulation.

More interestingly, the right panel of Figure~\ref{fig:Bbg} shows the scatter plot of the relative orientation between local and background magnetic fields as a function of the core mass, for both unbound (red) and bound (blue) cores. 
It appears that $B_{\rm core}$ and $B_{\rm bg}$ may be better aligned in more massive cores: almost all cores (both bound and unbound) with $M_{\rm core} \gtrsim 1 M_\odot$ have $\measuredangle[B_{\rm bg}, B_{\rm core}] \lesssim 30^\circ$. This could indicate that cores form within relatively quiescent environments (so that the local magnetic field is less perturbed compared to the background magnetic field) have higher chances to accrete more mass during these early stages.

Indeed, we note that though the majority of the cores have $B_{\rm core}$ roughly aligned with $B_{\rm bg}$, there are still about $\sim 10$\% of cores that have $\cos(\measuredangle[B_{\rm core}, B_{\rm bg}]) < 0.3$, or equivalently, $\measuredangle[B_{\rm core}, B_{\rm bg}] > 70^\circ$ (see Figure~\ref{fig:Bbg}).
Though the naive interpretation of
these non-alignments between core-scale and background magnetic fields is that these cores are more evolved (so that the steeper gravitational potential could be dominating over the magnetic tension force and twisting the local field direction relative to the mean magnetic field direction at larger scales), 
this is unlikely the case, because from the right panel of Figure~\ref{fig:Bbg} it is obvious that these cores are mostly less massive. In addition, the distribution of the relative orientation between core-scale and background magnetic fields does not seem to depend on whether or not the cores are gravitationally bound. 
Since the gravitationally bound cores are expected to be more evolved, this suggests that the misaligned core-scale and background magnetic fields are most likely due to local turbulence that is strong enough to alter the core-scale magnetic field morphology.

Of course, we cannot rule out the possibility that some cores with negative total energies are less evolved. 
In fact, as indicated in the right panel of Figures~\ref{fig:Bbg} which plots $\measuredangle[B_{\rm bg}, B_{\rm core}]$ vs.~core mass, these gravitationally bound cores are not necessarily the most massive ones either. 
Since the gas kinetic energy within the core (which could be either supporting the core against collapse or the direct result of core collapsing) is not included in calculating the total energy and determining the binding status, this could further suggest that the core-scale gas turbulent motion is strong enough in general to affect the progression of core evolution. 
This is consistent with the results discussed in \cite{2018ApJ...865...34C}, where up to $\sim 90\%$ of the kinetic energies within their simulated cores were found to be dominated by turbulent motions.

\subsubsection{Core orientation and core-scale magnetic field: the core$-$local magnetic field alignment}
\label{sec::simBcore}

\begin{figure*}
    \centering
    \includegraphics[width=\textwidth]{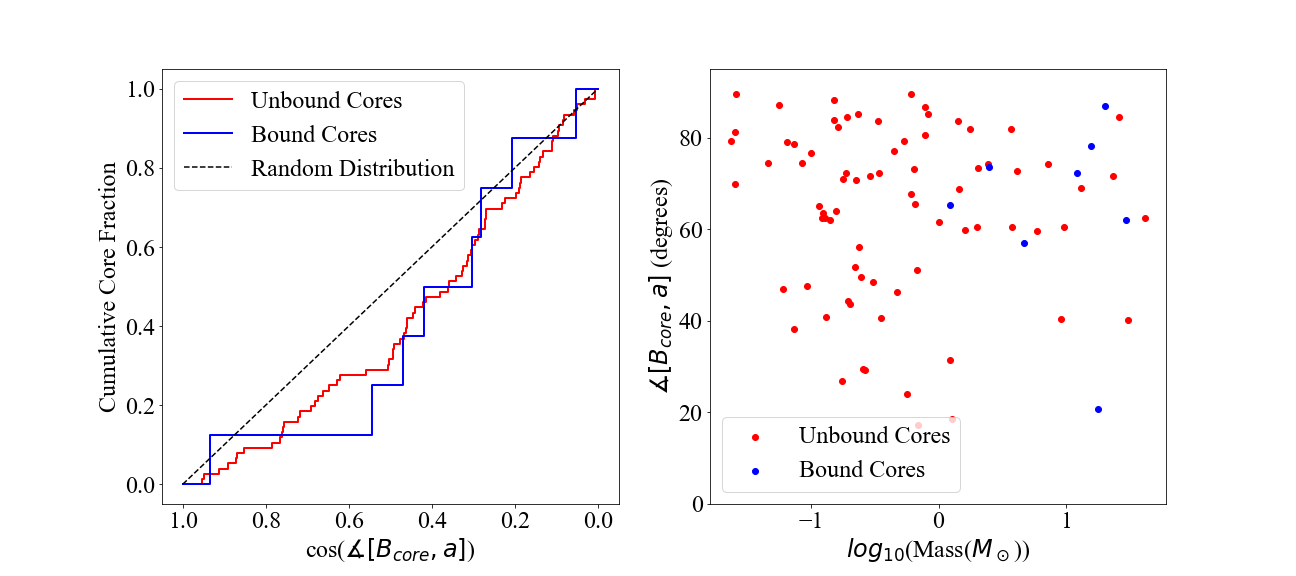}
    \caption{Same as Figure~\ref{fig:Bbg} but showing the angle between the core major axis and core-scale magnetic field, $\measuredangle[B_{\rm core}, a]$.}
    \label{fig:Bmajor}
\end{figure*}

Figure~\ref{fig:Bmajor} (left panel) shows the cumulative distribution (as a step function) of the relative angles (in cosine) between the major axes of cores and the core-scale magnetic field direction. 
When comparing with the random distribution, the K-S statistics is 0.21 ($p$-value $\approx 0.04$) and 0.33 ($p$-value $\approx 0.31$) for unbound and bound cores, respectively. 
These values are smaller than those for the local$-$background magnetic field alignment (see Section~\ref{sec::simBcBbg}) but still large enough to indicate that the core$-$local magnetic field alignment is not completely random, especially for the unbound cores with the $<5\%$ $p$-value.
Looking at the step function, the slightly more rapid rise at greater angles (smaller values of cosine) suggests that there is a weak but clear preference for dense cores to align perpendicular to the local magnetic fields. 
Given that it is easier for material to move along magnetic field lines, this tendency can be understood as the result of magnetized cores contracting more easily along the local magnetic field direction. Thus, the elongation of the cores tends to be perpendicular to the local magnetic field direction.
This is consistent with the result discussed in \cite{2018ApJ...865...34C}, where the roughly perpendicular core--local magnetic field alignments were observed in simulations with relatively stronger magnetization levels (see their Figure~6).

Similar to Figure~\ref{fig:Bbg}, we plot the relative orientation (in degree) between cores and local magnetic fields as a function of core mass in the right panel of Figure~\ref{fig:Bmajor}. Though there is no obvious trend with mass seen, one can still tell that most of the massive cores ($M_{\rm core} \gtrsim 1 M_\odot$) have angles $\gtrsim 60^\circ$ between their major axes and local magnetic fields. This indicates that more massive cores may be more likely to align perpendicular to the local magnetic field. 
These more massive cores could have accreted more material and contracted more severely along the local field lines and therefore show a stronger preference for perpendicular alignment.

\subsubsection{Core Orientation and Background Magnetic Field: the core$-$background magnetic field alignment}
\label{sec::simBcloud}

\begin{figure*}
    \centering
   \includegraphics[width=\textwidth]{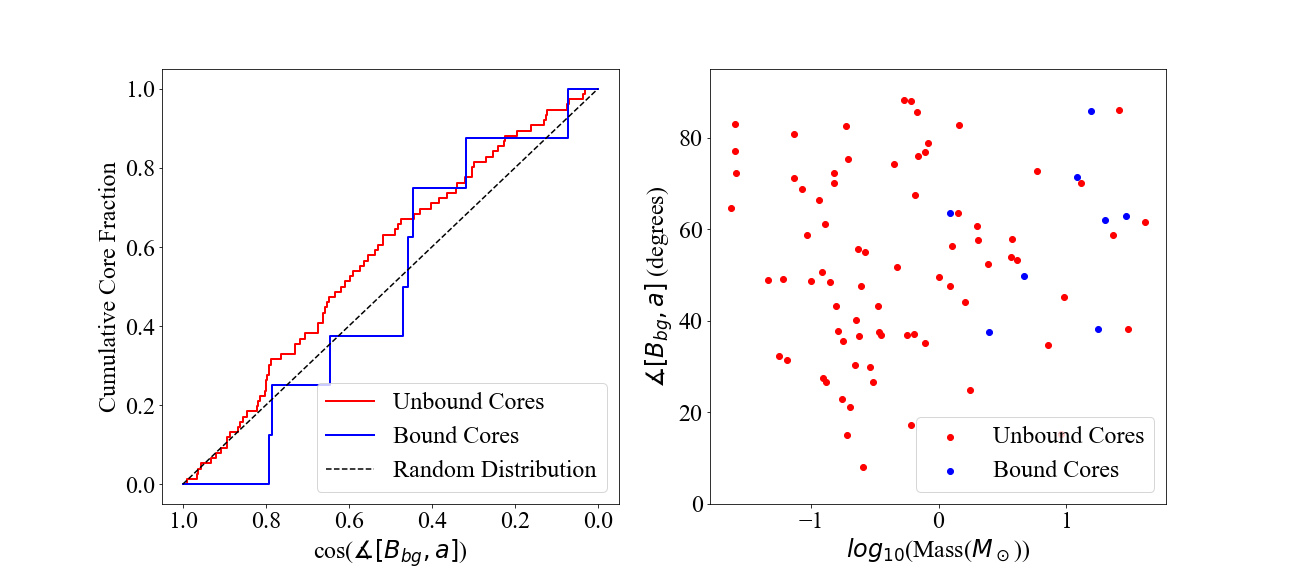}
    \caption{Same as Figure~\ref{fig:Bbg} but showing the angle between core major axis and background magnetic field, $\measuredangle[B_{\rm bg}, a]$.}
    \label{fig:Bgmajor}
\end{figure*}


Interestingly, 
Figure~\ref{fig:Bgmajor} (left panel) shows the distribution of angles between the background field (at the scale of 0.5~pc) and core major axis is very different from the $\measuredangle[B_{\rm core}, a]$-distribution shown in Figure~\ref{fig:Bmajor}, 
i.e.,~the core orientation seems to be slightly parallel, or even relatively random (especially for bound cores) with respect to the background magnetic field.
The K-S statistic relative to a random distribution (dashed diagonal line in the plot) gives 0.15 and 0.21 for unbound and bound cores, respectively (see Table~\ref{tab:cores}). Both are smaller than that of the relative alignment between core orientation and core-scale magnetic field (0.21 and 0.33; see Section~\ref{sec::simBcore}). 
In addition, the $p$-values (0.25 for unbound and 0.86 for bound cores; see Table~\ref{tab:cores}) are both relatively large, which suggests that the hypothesis of random distributions cannot be ruled out.

This indicates that, though core orientation is strongly affected by the local magnetic field within the core itself, the background magnetic field morphology in the pc-scale surroundings of the forming dense core has a relatively weak impact on the core-formation process.
This result likely reflects the fact that this simulated cloud where all these cores form is a turbulent medium with a moderately trans- to super-Alfv\'enic Mach number $\sim 1.5$, i.e.,~the larger-scale magnetic field is not strong enough to dominate the gas dynamics within the cloud. That being said, recall that the local and background magnetic field are in fact strongly correlated, as shown in Figure~\ref{fig:Bbg}. Looking back at Figure~\ref{fig:Bbg}, we see that the angle differences between $B_{\rm bg}$ and $B_{\rm core}$ for individual cores are mostly small ($\lesssim 30^\circ$), but 
this `noise' is large enough to wash out the weak preference of perpendicular alignment in $\measuredangle[B_{\rm core}, a]$ (mostly $\gtrsim 45^\circ$) and make $\measuredangle[B_{\rm bg}, a]$ more or less a random distribution.

We again plot the relative core$-$background magnetic field alignment as a function of core mass in the right panel of Figure~\ref{fig:Bgmajor}.
The distribution seems very random, with a similar, but weaker, feature as in Figure~\ref{fig:Bmajor}, that most of the massive cores ($M_{\rm core} \gtrsim 1 M_\odot$) have core$-$background magnetic field alignment larger than a certain value ($\sim 40^\circ$; $\sim 60^\circ$ for core$-$local magnetic field alignment).
Though the preference is not significant, 
this is consistent with our argument in the previous sections, that cores forming within relatively quiescent environment have better chances to grow and become more massive.

\subsection{Core$-$Background Magnetic Field Alignment in Synthetic Observations}
\label{sec::synobs}


\begin{figure}
    \centering
    \includegraphics[width=\columnwidth]{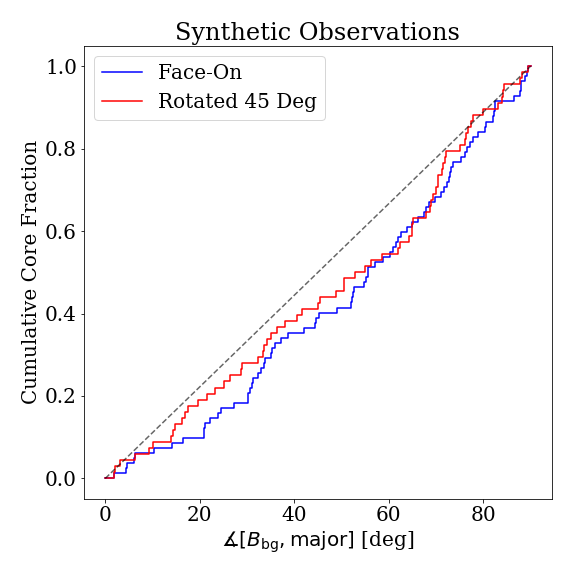}
    \caption{The step functions of the relative angle between the orientation of the major axes of the cores and the orientation of the pc-scale magnetic field near the core location ($B_{\rm bg}$) for synthetic observations along the face-on ({\it blue}) and rotated ({\it red}) projections. The dashed gray line shows the expected distribution for no preference in orientation. }
    \label{fig:sim_2d_step}
\end{figure}

To match the 15$^\prime$ resolution of {\it Planck} polarimetry,
the synthetic maps of Stokes $q$ and $u$ parameters and column density $N_{\rm H}$ (calculated from simulated data as described in Section~\ref{sec::methods}) were smoothed by convolving with 2D Gaussian kernels so that the final polarization maps have an equivalent beam size of 15$^\prime$ at a distance of $300$~pc (i.e.,~resolution $\sim 1$~pc).
We then use the locations of unsmoothed column density peaks within cores identified by \texttt{astrodendro} to calculate the inferred background magnetic field orientation at core locations, $\chi_{\rm core}$, from our smoothed $q$ and $u$ maps as defined in Equation~(\ref{eq:synchi}) for both the face-on and $45^\circ$-inclined synthetic observations. 
We can therefore calculate the relative alignment angle between $\chi_{\rm core}$ and the position angle of the major axis of the core, $|\chi_{\rm core} - {\rm PA}_{\rm major}|$, which falls within the range $[0, 90^\circ]$.
Note that the cores are typically much smaller than the resolution of the synthetic polarization maps, and thus this angle measurement is more comparable to the analysis in Section~\ref{sec::simBcloud} than that of Section~\ref{sec::simBcore}. We thus denote the angle  $|\chi_{\rm core} - {\rm PA}_{\rm major}|$ as $\measuredangle[B_{\rm bg}, a]$ to be consistent with our notation in the previous sections.

The results are summarized in Figure~\ref{fig:sim_2d_step}, which shows the cumulative distribution function (as a step function) of the relative angle between the major axes of cores identified in projected column densities and the polarization-indicated magnetic field direction at a $\sim 1$~pc scale, for both the face-on (blue line) and $45^\circ$-inclined (red line) cases. 
The results from the two cases are highly similar with almost the same values from the K-S statistics (0.17 and 0.15 for face-on and $45^\circ$ case, respectively), which indicates that projection effects do not dramatically change the apparent alignment, though the $p$-values of the K-S tests differ slightly between these two projections (0.15 and 0.33, respectively; see Table~\ref{tab:cores}). 
Nevertheless, this similarity in the core--background magnetic field alignment could be related to the fact that the identified cores are mostly aligned with filamentary structures (see Figure~\ref{fig::simcores}), and the orientations of filaments do not change significantly after rotation. The viewing angle could have a stronger impact on local polarization direction, but at pc-scale which we are considering here, the polarization direction remains similar. As a result, the relative alignment between dense cores and background magnetic field does not significantly depend on the viewing angle. 

Figure~\ref{fig:sim_2d_step} also suggests that cores have a slight tendency to align perpendicular to the background magnetic field in these synthetic observations, with a median value of relative angle $\measuredangle[B_{\rm bg}, {\rm major}] \approx 50^\circ$ for both projections (see Table~\ref{tab:cores}). Though this value is generally consistent with that measured in (unbound) cores defined in 3D, note that the trend in alignment is very different: our 3D measurements indicate that cores tend to align slightly parallel (unbound cores) or relatively random (bound cores) to the background magnetic field (see Figure~\ref{fig:Bgmajor}). 
This difference could be partly caused by the difference in core elongation for cores defined in 3D and 2D (see Table~\ref{tab:cores} for the median values of $b/a$), which is due to the fact that cores defined by gravitational potential (our 3D study) will be intrinsically more rounded, as we discussed in Section~\ref{sec::simCoresId}. Also, note that projected cores are located mostly within filaments, while filaments tend to align perpendicular to their local magnetic field. We will return to this point in the Discussion Section~\ref{sec::aspRatio} below.

\subsection{Core--Background Magnetic Field Alignment in Observations}
\label{sec::obscores}

\begin{figure}
    \centering
    \includegraphics[width=\columnwidth]{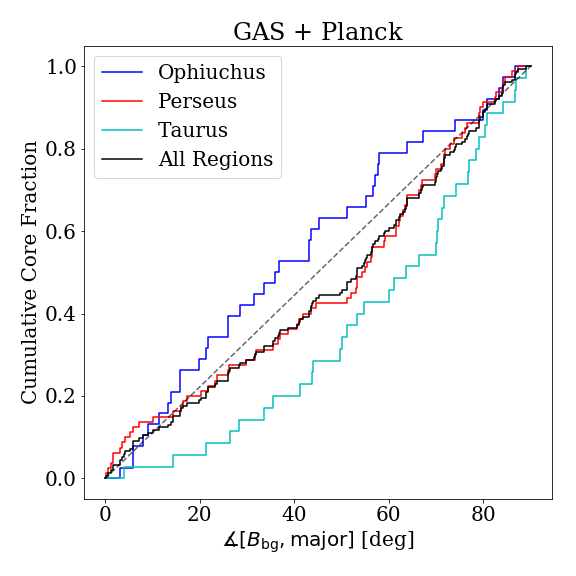}
    \caption{The step functions of the relative angle between the orientation of the major axes of the cores identified in GAS data, PA$_{\rm major}$, and the orientation of the polarization-inferred magnetic field from {\it Planck}, $B_{\rm POS}$, for Ophiuchus ({\it blue}), Perseus ({\it red}), Taurus ({\it cyan}), and all regions combined ({\it black}). The dashed gray line shows the expected distribution for no preference in orientation. }
    \label{fig:cloud_step}
\end{figure}

For the cores identified in \amm\ emission in the GAS dataset, we extract the Stokes parameters $Q$ and $U$ 
from the {\it Planck} data at the location of each to compute the inferred background magnetic field direction projected on the sky following Equation~(\ref{eqn:bangle}).
The relative alignment between dense cores and background magnetic fields is then derived following the same process discussed in Section~\ref{sec::synobs}.
The results from individual clouds are summarized in Figure~\ref{fig:cloud_step}, which shows cumulative distributions of the core$-$background magnetic field alignment angle. Results for individual regions within each cloud are included in Appendix~\ref{sec::GASmaps}.

We note that there are systematic differences among the three clouds examined in the relative alignment of their cores with respect to the pc-scale magnetic field. Namely, Taurus shows a preference for its cores to align perpendicular to the background magnetic field, rather than parallel to it. Perseus shows a relatively random distribution between its core major axes and the background magnetic field direction. Ophiuchus shows a preference for parallel alignment of its cores' long-axes and the background magnetic field. 
This difference is quantitatively reflected by the median values of the relative angles between background magnetic fields and the orientations of core major axes, $\langle\measuredangle[B_{\rm bg}, a]\rangle$, listed in Table~\ref{tab:cores},
which are $63.7^\circ$, $54.7^\circ$, and $36.3^\circ$ for Taurus, Perseus, and Ophiuchus, respectively. 
In addition, the K-S statistic (comparing to a random distribution) is 0.26 for Taurus with $p$-value less than $5\%$ (see Table~\ref{tab:cores}), which suggests that there is a clear preference of core$-$background magnetic field alignment in Taurus. 
In contrast, both Ophiuchus and Perseus have small K-S statistics with large $p$-values, indicating there is no evidence that the core$-$background magnetic field alignments in these two clouds are different from a random distribution.
We thus cannot rule out the possibility that dense cores in Ophiuchus and Perseus have no preferred alignment with respect to the background magnetic field.

We further note that there are regional differences across individual clouds in core$-$background magnetic field alignment. As listed in Table~\ref{tab:cores}, while the aspect ratios of identified cores remain similar across multiple regions within the cloud, the median value of the relative angles between core major axes and background magnetic fields ranges from $\approx 29^\circ$ in L1455 to $\approx 64^\circ$ in IC348 for Perseus, and from $\approx 44^\circ$ in B18 to $\approx 76^\circ$ in HC2 for Taurus.
The more than a factor of 2 difference in preferred alignment between core and pc-scale magnetic field within Perseus may be related to the fact that different regions within Perseus span a large range in star formation activities. 

Last but not the least, when comparing Figure~\ref{fig:cloud_step} with the results from the synthetic observations (Figure~\ref{fig:sim_2d_step}), our result suggests that Taurus and Perseus seem to match our specific simulation better than Ophiuchus. We include a more detailed discussion on the magnetic field structure in each cloud in Section~\ref{sec::cloudB}. 

\section{Discussion}
\label{sec::disc}

\subsection{Comparing Cores Identified in 2D and 3D}
\label{sec::aspRatio}

One goal of this study is to compare the properties of cores identified in 3D space and in projected 2D maps.
One can easily tell by looking at Figure~\ref{fig::simcores} that the locations of cores identified in 2D and 3D are similar, and the \texttt{astrodendro} method applied to the projected map does recover the bound cores.
The unbound 3D cores are sometimes recovered, mostly when they are on higher column density filaments,
though our dendrogram analysis seems to miss some of the 3D cores in low column density regions (see e.g., the region near $x=8$~pc, $y=2$~pc in the left and middle panels of Figure~\ref{fig::simcores}).
We also note that though the median values of the core sizes are similar for 2D and 3D cores (see Table~\ref{tab:cores}), some 3D cores are broken up into several smaller cores in 2D by the dendrogram algorithm, which suggests that there are sub-structures within the 3D cores that are highlighted after projection.
The gravitationally bound core at $x\approx 9$~pc, $y\approx 8$~pc is one example for this scenario (see the left and middle panels of Figure~\ref{fig::simcores}).
These results are consistent to that discussed in \cite{Beaumont+13}, which suggested that the observed intensity features do not always correspond to the real gas structures.


More importantly, we point out that in our analysis, 2D cores tend to be more elongated than 3D cores (see Table~\ref{tab:cores}).
In addition to the intrinsic difference between identifying cores based on gravitational potential and gas emission that we discussed in Section~\ref{sec::coreId}, the core--filament correlation also plays a critical role here.
In general, we observe that cores tend to be elongated in the direction parallel to their filaments.  This tendency is clearly seen for the cores extracted with \texttt{astrodendro} in the synthetic observations shown in the center and right panels of Figure~\ref{fig::simcores}. It is also apparent, however, in some identified cores within the simulation in 3D, particularly the bound cores, as shown in the left panel of Figure~\ref{fig::simcores}.
In the \amm{} observations by GAS, we also see some cores preferentially aligned parallel to their host filaments (see Figures~\ref{fig:mapOph}$-$\ref{fig:mapTau}).  This behavior is not unexpected given that previous studies already noted that dense cores are generally associated with filaments \citep{Konyves+10,adw+14}.
Nevertheless, considering the slightly inconsistent results in relative alignment between cores and pc-scale magnetic fields measured in 3D and 2D (Figures~\ref{fig:Bgmajor} and \ref{fig:sim_2d_step}), our study suggests that the shape and orientation of dense cores could be very different after being projected along the line of sight.

\subsection{Magnetic Field Structures in MCs}
\label{sec::cloudB}

As mentioned in Section~\ref{sec::obscores}, the difference in core$-$background magnetic field alignment could provide clues about the relative energetic importance of magnetic fields at cloud scale, compared to turbulent gas motions or feedback. In general, clouds with stronger magnetic fields should have core major axes aligned more perpendicular to the field, because it is much easier to contract or collapse along the magnetic field lines than across them. 
Similarly, past studies have also shown that on scales of $\sim 1-5$~pc, dense, star-forming filaments tend to align perpendicular to the surrounding magnetic field \citep{adw+14,caa16}. 
Our results from the synthetic observations are also consistent with this picture, showing a slight preference of perpendicular alignment between cores and the pc-scale magnetic field, within a trans- to super-Alfv\'enic simulation. 

Detailed studies of {\it Planck} maps of nearby star-forming MCs have previously shown that high column density structures often tend to have a preference for alignment perpendicular to the magnetic field, while low column density structures preferentially align parallel to the magnetic field \citep{pxxxv16,Jow+18,Soler+19}. This change in relative orientation within the cloud has been interpreted as the signature of dynamically important magnetic fields, and an indication that gas accumulation onto dense filaments and cores is preferentially occurring in the direction parallel to the magnetic field. 

Indeed, in \cite{pxxxv16} the degree of alignment between gas structure and magnetic field varies significantly from cloud to cloud, similar to what we see in the core$-$background magnetic field relative alignments in Perseus, Taurus, and Ophiuchus (Figure~\ref{fig:cloud_step}). 
Though the absolute value of magnetic field strength cannot be directly measured, the relative significance of large-scale magnetic field can be inferred by investigating the polarization morphology within the cloud. Clouds with relatively strong magnetic fields that dominate the gas dynamics tend to have more well-ordered polarization patterns, which correspond to lower values of the dispersion of polarization angles, ${\cal S}$, defined as the averaged difference in the polarization angles for all points within a specified lag scale around each pixel \citep{planckXIX}:
\begin{equation}\label{eq:S}
    \mathcal{S}(x,\delta) = \sqrt{\frac{1}{N}\sum_\text{i=1}^N(\Delta\psi_{x,i})^2},
\end{equation}
where $\delta$~is the lag scale and
\begin{equation}\label{eq:psi}
    \Delta\psi_{x,i} = \psi_x - \psi_i
\end{equation}
is the difference in polarization angle between a given map location $x$ and a nearby map location $i$ at a distance of $\delta$.
A recent study of the {\it Planck} polarization maps by  \cite{Colin+19} shows that, at 15\arcmin{} scale, Taurus has a relatively small dispersion with ${\cal S} \approx 5.5^\circ$, while Perseus has ${\cal S}\approx 9.7^\circ$.\footnote{The standard deviation of ${\cal S}$ is usually measured in $\log$ space as $\sigma_{\log {\cal S}}$, and is $\approx 0.25$ for both Taurus and Perseus. }
This difference may be consistent with our arguments that the difference in the core$-$background magnetic field alignment probably depends on the relative magnetic strength and the magnetic field structure in the cloud: clouds with smaller ${\cal S}$ values tend to have cores that align more perpendicular to the background magnetic fields.

Such differences in ${\cal S}$ values as well as the relative core$-$background magnetic field alignment could also reflect the star formation activity in each cloud.
Taurus is a low mass and relatively diffuse MC with a relatively low star-formation rate
\citep[e.g.,][]{Onishi96,Goldsmith08}, and the {\it Planck} polarization map of Taurus shows a very ordered large-scale magnetic field \citep{planckXIX}. 
These observations suggest that Taurus may be relatively less perturbed comparing to other nearby star-forming regions.
Taurus also shows a strong tendency for dense filaments to align perpendicular to the ambient magnetic field \citep{adw+14,pxxxv16}. Since our identified cores tend to be elongated in the direction of the dense filaments (see Figure~\ref{fig:mapTau}), it is expected that we see a trend that the cores on average also have a preference to be aligned perpendicular to the pc-scale magnetic field.  
This tendency is consistent with (but stronger than) what we found in our synthetic observations (Figure~\ref{fig:sim_2d_step}), which also show tight core--filament correlations (see Figure~\ref{fig::simcores}).

Perseus spans a range in star formation activities, from smaller regions that are less active like Taurus, to active star-forming regions like NGC\,1333 \citep[e.g.,][]{Enoch+06}.
Perseus also has a much more disordered magnetic field: as mentioned above, \cite{Colin+19} shows that the geometric mean of the disorder in the magnetic field on 15$^\prime$~scales (${\cal S}$) is $9.7^\circ$ for Perseus, considerably larger than the ${\cal S} = 5.5^\circ$ for Taurus.
In addition to the possible field distortion from several expanding shells/bubbles observed in CO lines \citep{Arce+11},
\cite{Colin+19} argue that at least part of the large ${\cal S}$ values for Perseus are due to projection effects.  Using a method of deriving mean cloud magnetic field inclination angles from polarization observations first presented in \cite{2019MNRAS.485.3499C}, they estimate an inclination angle of $69^\circ$, the highest in their sample of 9 nearby MCs.  This large inclination could partially explain why the projected magnetic field appears so disordered in Perseus.  In addition, many of the regions in Perseus, such as NGC\,1333, show complicated filamentary structures on small scales without a clearly preferred orientation (see Figure~\ref{fig:mapPer}), and therefore these regions appear to have no preferential alignment between the cores and the large-scale magnetic field traced by {\it Planck}.
Since most regions in Perseus do not have prominent large-scale filamentary structures like HC2 or B211/B213 in Taurus (see Figures~\ref{fig:mapPer} and \ref{fig:mapTau}), it is not surprising that the core--background magnetic field alignment in Perseus appears to be slightly more random (see the K-S statistics and $p$-values in Table~\ref{tab:cores}) than that measured in our synthetic observations, where the identified core shapes and locations are strongly correlated with the high-column density filaments (see Figure~\ref{fig::simcores}).
Our results are also consistent with the scenario suggested by \cite{Stephens+17}, who argued that the core-scale dynamics may be independent of the large-scale structure in Perseus, because a random distribution between protostellar outflows and filamentary structures in Perseus cannot be ruled out.

The third cloud, Ophiuchus, is a very active star forming region. 
The Ophiuchus regions have higher column densities, are the most actively star-forming, and are warmer by a few K
\citep{1992lmsf.book..159W}.
Ophiuchus also shows a very different magnetic field morphology, where the high column density structure has very little preference for aligning either parallel or perpendicular to the {\it Planck}-scale magnetic field \citep[see, e.g.,][]{pxxxv16,Jow+18}. 
This could indicate the existence of super-Alfv\'enic turbulence.
Indeed, the maps of the inferred magnetic field shown in \cite{pxxxv16} show considerable curvature in the large-scale magnetic field.  
It has been proposed that parts of the cloud have been compressed by expanding shells associated with one or more supernova explosions \citep{2008hsf2.book..351W},
which makes interpretation of polarization patterns slightly different. 
One possible explanation for the magnetic field geometry is that the magnetic field has been altered by feedback, so that in some regions the magnetic field is closer to aligning parallel to the dense filaments,\footnote{Such parallel alignment between magnetic field and dense filament has also been reported in the OMC1 region of the Orion A molecular cloud by \cite{Monsch+18}.} and therefore also preferentially parallel to the dense cores within the filaments.
This could explain the obvious discrepancy on the core--background magnetic field alignment between Ophiuchus (slightly parallel; Figure~\ref{fig:cloud_step}) and the synthetic observations investigated in Section~\ref{sec::synobs} (slightly perpendicular; Figure~\ref{fig:sim_2d_step}), because protostellar feedback like outflows are not included in the simulation discussed in this study, which might be the key mechanism affecting the core orientations and filamentary structures in these active star-forming regions,
and may also provide insight into the importance of the magnetic field \citep[see e.g.,][]{LeeHull+17}.

While further analysis involving more star-forming MCs and simulations with various levels of magnetization will better quantify this correlation, our results clearly suggest that there is a preference on the relative alignment between dense cores and background magnetic field that depends on the magnetized environment within the cloud. 
We would also like to point out that since GAS data contains velocity dispersion information, we will be able to derive the turbulence levels within individual clouds and bring gas dynamics into the picture in our follow-up studies.

\section{Summary and Conclusion}
\label{sec::sum}

In this study, we investigated the relative alignment between the orientations of dense cores and magnetic field direction at various scales within star-forming MCs using both simulation and observational data. We found that the core$-$background magnetic field alignment could depend on the magnetization level within the cloud, though projection from 3D to 2D may also have an impact on the measured alignment. Our main conclusions are summarized below:
\begin{itemize}
    \item Utilizing a star-forming, MC-scale MHD simulation, we studied the correlation between local (core-scale, $\sim 0.01$~pc) and background (pc-scale, $0.5$~pc) magnetic field directions. We found that the local field is generally aligned with the large-scale field (approximately 50\% of cores have $\measuredangle[B_{\rm bg}, B_{\rm core}]<30^\circ$; see Figure~\ref{fig:Bbg}), though there are still a good fraction of cores with very different local field direction (approximately 10\% of cores have $\measuredangle[B_{\rm bg}, B_{\rm core}]>70^\circ$). These cores with large differences in core- and pc-scale magnetic field are likely more turbulent (see Section~\ref{sec::simBcBbg}).
    
    \item We found that in 3D space, cores have a weak but clear preference to be aligned perpendicular to the local magnetic field, while being oriented slightly parallel or completely random to the background magnetic field (Figures~\ref{fig:Bmajor} and \ref{fig:Bgmajor}). This tendency can be explained by anisotropic gas flows along magnetic field lines during core formation and contraction, which has been extensively discussed in \cite{2014ApJ...785...69C,2015ApJ...810..126C,2018ApJ...865...34C}.
    
    \item By comparing cores defined in 3D with those identified in projected 2D maps, it is clear that 2D cores are mostly associated with filaments, and thus are more elongated than 3D cores (Figure~\ref{fig::simcores}). 
     Though the intrinsic difference between the gravitational potential morphology and real gas distribution cannot be ignored (which makes the identified 3D cores more rounded; see Section~\ref{sec::simCoresId}), the close alignment between 2D projected cores and filaments implies that the core extraction algorithm sees the cores as part of filaments and therefore being longer, while the underlying structures (which could be more rounded based on the identified 3D cores) remain unseen.
    This may explain the difference in core$-$background magnetic field alignment we see in the simulation (relatively random; Figure~\ref{fig:Bgmajor}) and synthetic observations (preferrably perpendicular; Figure~\ref{fig:sim_2d_step}). 
    Nevertheless, our analysis suggests that care needs to be made when interpreting such alignments from observations.
    
    \item There is a systematic difference in the core$-$background magnetic field alignment among the three nearby MCs that we investigated (Figure~\ref{fig:cloud_step}), with the median value of the relative angle between core major axis and background magnetic field varies from 36.3$^\circ$ (Ophiuchus), $54.7^\circ$ (Perseus), to $63.7^\circ$ (Taurus).
    Looking at the cumulative distributions (Figure~\ref{fig:cloud_step}) and combining with the K-S statistics (see Table~\ref{tab:cores}), these clouds have 2D core orientations that are either slightly more perpendicular (Taurus), slightly more random (Perseus), or significantly more parallel (Ophiuchus) to the background magnetic field than that observed in the synthetic observations (slight preference of perpendicular alignment; Figure~\ref{fig:sim_2d_step}), which were generated from a trans- to super-Alfv\'enic simulation.
    Combined with other polarimetric properties, we suggest that the relative alignment between dense cores and pc-scale magnetic fields depends on the magnetic properties of the cloud, and could potentially provide an alternative approach than the traditional Zeeman observations on investigating the magnetization level of the cloud.
    
\end{itemize}

\section*{Acknowledgements}
The authors would like to thank Juan Soler, who originally provided the smoothed {\it Planck} polarization maps,
and Mark Heyer for encouraging critiques that improved the paper. 
C-YC, LMF, and Z-YL acknowledge support from NSF grant AST-1815784. 
EAB was supported by a REU summer research fellowship at the National Radio Astronomy Observatory (NRAO), and LMF acknowledges support as a Jansky Fellow of NRAO. 
NRAO is a facility of the National Science Foundation (NSF operated under cooperative agreement by Associated Universities, Inc.).
LMF and Z-YL acknowledge support from NASA~80NSSC18K0481. Z-YL is supported in part by NASA~80NSSC18K1095 and NSF AST-1716259.
AP acknowledges the financial support of the the Russian Science Foundation project 19-72-00064.
SSRO and HHC acknowledge support from a Cottrell Scholar Award from Research Corporation.
AC-T acknowledges support from MINECO project AYA2016-79006-P.
This research made use of {\tt astropy} \citep{apy13,apy18} and {\tt astrodendro}, a {\tt Python} package to compute dendrograms of astronomical data. The authors thank the staff at the Green Bank Telescope for their help facilitating the Green Bank Ammonia Survey. The Green Bank Observatory is a facility of the National Science Foundation operated under cooperative agreement by Associated Universities, Inc. 

\bibliographystyle{aasjournal}
\bibliography{reference}

\begin{thebibliography}{}
\expandafter\ifx\csname natexlab\endcsname\relax\def\natexlab#1{#1}\fi

\bibitem[{{Andr{\'e}} {et~al.}(2009){Andr{\'e}}, {Basu}, \&
  {Inutsuka}}]{Andre2009}
{Andr{\'e}}, P., {Basu}, S., \& {Inutsuka}, S. 2009, {The formation and
  evolution of prestellar cores}, ed. G.~{Chabrier}, 254

\bibitem[{{Andr{\'e}} {et~al.}(2014){Andr{\'e}}, {Di Francesco},
  {Ward-Thompson}, {Inutsuka}, {Pudritz}, \& {Pineda}}]{adw+14}
{Andr{\'e}}, P., {Di Francesco}, J., {Ward-Thompson}, D., {et~al.} 2014, in
  Protostars and Planets VI, ed. H.~{Beuther}, R.~S. {Klessen}, C.~P.
  {Dullemond}, \& T.~{Henning}, 27

\bibitem[{{Arce} {et~al.}(2011){Arce}, {Borkin}, {Goodman}, {Pineda}, \&
  {Beaumont}}]{Arce+11}
{Arce}, H.~G., {Borkin}, M.~A., {Goodman}, A.~A., {Pineda}, J.~E., \&
  {Beaumont}, C.~N. 2011, \apj, 742, 105

\bibitem[{{Astropy Collaboration} {et~al.}(2013){Astropy Collaboration},
  {Robitaille}, {Tollerud}, {Greenfield}, {Droettboom}, {Bray}, {Aldcroft},
  {Davis}, {Ginsburg}, {Price-Whelan}, {Kerzendorf}, {Conley}, {Crighton},
  {Barbary}, {Muna}, {Ferguson}, {Grollier}, {Parikh}, {Nair}, {Unther},
  {Deil}, {Woillez}, {Conseil}, {Kramer}, {Turner}, {Singer}, {Fox}, {Weaver},
  {Zabalza}, {Edwards}, {Azalee Bostroem}, {Burke}, {Casey}, {Crawford},
  {Dencheva}, {Ely}, {Jenness}, {Labrie}, {Lim}, {Pierfederici}, {Pontzen},
  {Ptak}, {Refsdal}, {Servillat}, \& {Streicher}}]{apy13}
{Astropy Collaboration}, {Robitaille}, T.~P., {Tollerud}, E.~J., {et~al.} 2013,
  \aap, 558, A33

\bibitem[{{Astropy Collaboration} {et~al.}(2018){Astropy Collaboration},
  {Price-Whelan}, {Sip{\H o}cz}, {G{\"u}nther}, {Lim}, {Crawford}, {Conseil},
  {Shupe}, {Craig}, {Dencheva}, {Ginsburg}, {VanderPlas}, {Bradley},
  {P{\'e}rez-Su{\'a}rez}, {de Val-Borro}, {Aldcroft}, {Cruz}, {Robitaille},
  {Tollerud}, {Ardelean}, {Babej}, {Bach}, {Bachetti}, {Bakanov}, {Bamford},
  {Barentsen}, {Barmby}, {Baumbach}, {Berry}, {Biscani}, {Boquien}, {Bostroem},
  {Bouma}, {Brammer}, {Bray}, {Breytenbach}, {Buddelmeijer}, {Burke},
  {Calderone}, {Cano Rodr{\'{\i}}guez}, {Cara}, {Cardoso}, {Cheedella},
  {Copin}, {Corrales}, {Crichton}, {D'Avella}, {Deil}, {Depagne}, {Dietrich},
  {Donath}, {Droettboom}, {Earl}, {Erben}, {Fabbro}, {Ferreira}, {Finethy},
  {Fox}, {Garrison}, {Gibbons}, {Goldstein}, {Gommers}, {Greco}, {Greenfield},
  {Groener}, {Grollier}, {Hagen}, {Hirst}, {Homeier}, {Horton}, {Hosseinzadeh},
  {Hu}, {Hunkeler}, {Ivezi{\'c}}, {Jain}, {Jenness}, {Kanarek}, {Kendrew},
  {Kern}, {Kerzendorf}, {Khvalko}, {King}, {Kirkby}, {Kulkarni}, {Kumar},
  {Lee}, {Lenz}, {Littlefair}, {Ma}, {Macleod}, {Mastropietro}, {McCully},
  {Montagnac}, {Morris}, {Mueller}, {Mumford}, {Muna}, {Murphy}, {Nelson},
  {Nguyen}, {Ninan}, {N{\"o}the}, {Ogaz}, {Oh}, {Parejko}, {Parley}, {Pascual},
  {Patil}, {Patil}, {Plunkett}, {Prochaska}, {Rastogi}, {Reddy Janga},
  {Sabater}, {Sakurikar}, {Seifert}, {Sherbert}, {Sherwood-Taylor}, {Shih},
  {Sick}, {Silbiger}, {Singanamalla}, {Singer}, {Sladen}, {Sooley},
  {Sornarajah}, {Streicher}, {Teuben}, {Thomas}, {Tremblay}, {Turner},
  {Terr{\'o}n}, {van Kerkwijk}, {de la Vega}, {Watkins}, {Weaver}, {Whitmore},
  {Woillez}, {Zabalza}, \& {Astropy Contributors}}]{apy18}
{Astropy Collaboration}, {Price-Whelan}, A.~M., {Sip{\H o}cz}, B.~M., {et~al.}
  2018, \aj, 156, 123

\bibitem[{{Ballesteros-Paredes} {et~al.}(2007){Ballesteros-Paredes}, {Klessen},
  {Mac Low}, \& {Vazquez-Semadeni}}]{BP2007PPV}
{Ballesteros-Paredes}, J., {Klessen}, R.~S., {Mac Low}, M.~M., \&
  {Vazquez-Semadeni}, E. 2007, in Protostars and Planets V, ed. B.~{Reipurth},
  D.~{Jewitt}, \& K.~{Keil}, 63

\bibitem[{{Banerjee} {et~al.}(2009){Banerjee}, {V{\'a}zquez-Semadeni},
  {Hennebelle}, \& {Klessen}}]{2009MNRAS.398.1082B}
{Banerjee}, R., {V{\'a}zquez-Semadeni}, E., {Hennebelle}, P., \& {Klessen},
  R.~S. 2009, \mnras, 398, 1082

\bibitem[{{Basu} \& {Ciolek}(2004)}]{Basu+04}
{Basu}, S., \& {Ciolek}, G.~E. 2004, \apjl, 607, L39

\bibitem[{{Beaumont} {et~al.}(2013){Beaumont}, {Offner}, {Shetty}, {Glover}, \&
  {Goodman}}]{Beaumont+13}
{Beaumont}, C.~N., {Offner}, S. S.~R., {Shetty}, R., {Glover}, S. C.~O., \&
  {Goodman}, A.~A. 2013, \apj, 777, 173

\bibitem[{{Chen} {et~al.}(2016){Chen}, {King}, \& {Li}}]{2016ApJ...829...84C}
{Chen}, C.-Y., {King}, P.~K., \& {Li}, Z.-Y. 2016, \apj, 829, 84

\bibitem[{{Chen} {et~al.}(2019{\natexlab{a}}){Chen}, {King}, {Li}, {Fissel}, \&
  {Mazzei}}]{2019MNRAS.485.3499C}
{Chen}, C.-Y., {King}, P.~K., {Li}, Z.-Y., {Fissel}, L.~M., \& {Mazzei}, R.~R.
  2019{\natexlab{a}}, \mnras, 485, 3499

\bibitem[{{Chen} {et~al.}(2017){Chen}, {Li}, {King}, \&
  {Fissel}}]{2017ApJ...847..140C}
{Chen}, C.-Y., {Li}, Z.-Y., {King}, P.~K., \& {Fissel}, L.~M. 2017, \apj, 847,
  140

\bibitem[{{Chen} \& {Ostriker}(2014)}]{2014ApJ...785...69C}
{Chen}, C.-Y., \& {Ostriker}, E.~C. 2014, \apj, 785, 69

\bibitem[{{Chen} \& {Ostriker}(2015)}]{2015ApJ...810..126C}
---. 2015, \apj, 810, 126

\bibitem[{{Chen} \& {Ostriker}(2018)}]{2018ApJ...865...34C}
---. 2018, \apj, 865, 34

\bibitem[{{Chen} {et~al.}(2019{\natexlab{b}}){Chen}, {Storm}, {Li}, {Mundy},
  {Frayer}, {Li}, {Church}, {Friesen}, {Harris}, {Looney}, {Offner},
  {Ostriker}, {Pineda}, {Tobin}, \& {Chen}}]{GBTcore19}
{Chen}, C.-Y., {Storm}, S., {Li}, Z.-Y., {et~al.} 2019{\natexlab{b}}, \mnras,
  490, 527

\bibitem[{{Chen} {et~al.}(2019{\natexlab{c}}){Chen}, {Pineda}, {Goodman},
  {Burkert}, {Offner}, {Friesen}, {Myers}, {Alves}, {Arce}, {Caselli},
  {Chac{\'o}n-Tanarro}, {Chen}, {Di Francesco}, {Ginsburg}, {Keown}, {Kirk},
  {Martin}, {Matzner}, {Punanova}, {Redaelli}, {Rosolowsky}, {Scibelli}, {Seo},
  {Shirley}, {Singh}, \& {GAS Collaboration}}]{HopeGAS19}
{Chen}, H. H.-H., {Pineda}, J.~E., {Goodman}, A.~A., {et~al.}
  2019{\natexlab{c}}, \apj, 877, 93

\bibitem[{{Cox} {et~al.}(2015){Cox}, {Harris}, {Looney}, {Segura-Cox}, {Tobin},
  {Li}, {Tychoniec}, {Chandler}, {Dunham}, {Kratter}, {Melis}, {Perez}, \&
  {Sadavoy}}]{Cox15}
{Cox}, E.~G., {Harris}, R.~J., {Looney}, L.~W., {et~al.} 2015, \apjl, 814, L28

\bibitem[{{Cox} {et~al.}(2016){Cox}, {Arzoumanian}, {Andr{\'e}}, {Rygl},
  {Prusti}, {Men'shchikov}, {Royer}, {K{\'o}sp{\'a}l}, {Palmeirim}, {Ribas},
  {K{\"o}nyves}, {Bernard}, {Schneider}, {Bontemps}, {Merin}, {Vavrek}, {Alves
  de Oliveira}, {Didelon}, {Pilbratt}, \& {Waelkens}}]{caa16}
{Cox}, N.~L.~J., {Arzoumanian}, D., {Andr{\'e}}, P., {et~al.} 2016, \aap, 590,
  A110

\bibitem[{{Davis} \& {Greenstein}(1951)}]{DG1951}
{Davis}, Leverett, J., \& {Greenstein}, J.~L. 1951, \apj, 114, 206

\bibitem[{{Enoch} {et~al.}(2006){Enoch}, {Young}, {Glenn}, {Evans}, {Golwala},
  {Sargent}, {Harvey}, {Aguirre}, {Goldin}, {Haig}, {Huard}, {Lange},
  {Laurent}, {Maloney}, {Mauskopf}, {Rossinot}, \& {Sayers}}]{Enoch+06}
{Enoch}, M.~L., {Young}, K.~E., {Glenn}, J., {et~al.} 2006, \apj, 638, 293

\bibitem[{{Fiege} \& {Pudritz}(2000)}]{2000ApJ...544..830F}
{Fiege}, J.~D., \& {Pudritz}, R.~E. 2000, \apj, 544, 830

\bibitem[{{Fissel} {et~al.}(2016){Fissel}, {Ade}, {Angil{\`e}}, {Ashton},
  {Benton}, {Devlin}, {Dober}, {Fukui}, {Galitzki}, {Gandilo}, {Klein},
  {Korotkov}, {Li}, {Martin}, {Matthews}, {Moncelsi}, {Nakamura},
  {Netterfield}, {Novak}, {Pascale}, {Poidevin}, {Santos}, {Savini}, {Scott},
  {Shariff}, {Diego Soler}, {Thomas}, {Tucker}, {Tucker}, \&
  {Ward-Thompson}}]{2016ApJ...824..134F}
{Fissel}, L.~M., {Ade}, P. A.~R., {Angil{\`e}}, F.~E., {et~al.} 2016, \apj,
  824, 134

\bibitem[{{Fissel} {et~al.}(2019){Fissel}, {Ade}, {Angil{\`e}}, {Ashton},
  {Benton}, {Chen}, {Cunningham}, {Devlin}, {Dober}, {Friesen}, {Fukui},
  {Galitzki}, {Gandilo}, {Goodman}, {Green}, {Jones}, {Klein}, {King},
  {Korotkov}, {Li}, {Lowe}, {Martin}, {Matthews}, {Moncelsi}, {Nakamura},
  {Netterfield}, {Newmark}, {Novak}, {Pascale}, {Poidevin}, {Santos}, {Savini},
  {Scott}, {Shariff}, {Soler}, {Thomas}, {Tucker}, {Tucker}, {Ward-Thompson},
  \& {Zucker}}]{Fissel+19}
---. 2019, \apj, 878, 110

\bibitem[{{Foster} {et~al.}(2009){Foster}, {Rosolowsky}, {Kauffmann}, {Pineda},
  {Borkin}, {Caselli}, {Myers}, \& {Goodman}}]{Foster+09}
{Foster}, J.~B., {Rosolowsky}, E.~W., {Kauffmann}, J., {et~al.} 2009, \apj,
  696, 298

\bibitem[{{Franco} \& {Alves}(2015)}]{2015ApJ...807....5F}
{Franco}, G.~A.~P., \& {Alves}, F.~O. 2015, \apj, 807, 5

\bibitem[{{Friesen} {et~al.}(2017){Friesen}, {Pineda}, {co-PIs}, {Rosolowsky},
  {Alves}, {Chac{\'o}n-Tanarro}, {How-Huan Chen}, {Chun-Yuan Chen}, {Di
  Francesco}, {Keown}, {Kirk}, {Punanova}, {Seo}, {Shirley}, {Ginsburg},
  {Hall}, {Offner}, {Singh}, {Arce}, {Caselli}, {Goodman}, {Martin}, {Matzner},
  {Myers}, {Redaelli}, \& {GAS Collaboration}}]{fp+17}
{Friesen}, R.~K., {Pineda}, J.~E., {co-PIs}, {et~al.} 2017, \apj, 843, 63

\bibitem[{{Goldsmith} {et~al.}(2008){Goldsmith}, {Heyer}, {Narayanan}, {Snell},
  {Li}, \& {Brunt}}]{Goldsmith08}
{Goldsmith}, P.~F., {Heyer}, M., {Narayanan}, G., {et~al.} 2008, \apj, 680, 428

\bibitem[{{G{\'o}mez} \& {V{\'a}zquez-Semadeni}(2014)}]{2014ApJ...791..124G}
{G{\'o}mez}, G.~C., \& {V{\'a}zquez-Semadeni}, E. 2014, \apj, 791, 124

\bibitem[{{Gong} \& {Ostriker}(2011)}]{2011ApJ...729..120G}
{Gong}, H., \& {Ostriker}, E.~C. 2011, \apj, 729, 120

\bibitem[{{Gong} \& {Ostriker}(2015)}]{GO15}
{Gong}, M., \& {Ostriker}, E.~C. 2015, \apj, 806, 31

\bibitem[{{Heiles} {et~al.}(1993){Heiles}, {Goodman}, {McKee}, \&
  {Zweibel}}]{Heiles1993PP}
{Heiles}, C., {Goodman}, A.~A., {McKee}, C.~F., \& {Zweibel}, E.~G. 1993, in
  Protostars and Planets III, ed. E.~H. {Levy} \& J.~I. {Lunine}, 279

\bibitem[{{Heitsch} {et~al.}(2008){Heitsch}, {Hartmann}, {Slyz}, {Devriendt},
  \& {Burkert}}]{2008ApJ...674..316H}
{Heitsch}, F., {Hartmann}, L.~W., {Slyz}, A.~D., {Devriendt}, J. E.~G., \&
  {Burkert}, A. 2008, \apj, 674, 316

\bibitem[{{Heyer} {et~al.}(2008){Heyer}, {Gong}, {Ostriker}, \&
  {Brunt}}]{Heyer+08}
{Heyer}, M., {Gong}, H., {Ostriker}, E., \& {Brunt}, C. 2008, \apj, 680, 420

\bibitem[{{Hildebrand} {et~al.}(1984){Hildebrand}, {Dragovan}, \&
  {Novak}}]{1984ApJ...284L..51H}
{Hildebrand}, R.~H., {Dragovan}, M., \& {Novak}, G. 1984, \apjl, 284, L51

\bibitem[{{Hull} {et~al.}(2014){Hull}, {Plambeck}, {Kwon}, {Bower},
  {Carpenter}, {Crutcher}, {Fiege}, {Franzmann}, {Hakobian}, {Heiles}, {Houde},
  {Hughes}, {Lamb}, {Looney}, {Marrone}, {Matthews}, {Pillai}, {Pound},
  {Rahman}, {Sandell}, {Stephens}, {Tobin}, {Vaillancourt}, {Volgenau}, \&
  {Wright}}]{Hull+14}
{Hull}, C. L.~H., {Plambeck}, R.~L., {Kwon}, W., {et~al.} 2014, \apjs, 213, 13

\bibitem[{{Ikeda} {et~al.}(2009){Ikeda}, {Kitamura}, \&
  {Sunada}}]{2009ApJ...691.1560I}
{Ikeda}, N., {Kitamura}, Y., \& {Sunada}, K. 2009, \apj, 691, 1560

\bibitem[{{Inoue} \& {Fukui}(2013)}]{2013ApJ...774L..31I}
{Inoue}, T., \& {Fukui}, Y. 2013, \apjl, 774, L31

\bibitem[{{Johnstone} \& {Bally}(2006)}]{Johnstone06}
{Johnstone}, D., \& {Bally}, J. 2006, \apj, 653, 383

\bibitem[{{Jow} {et~al.}(2018){Jow}, {Hill}, {Scott}, {Soler}, {Martin},
  {Devlin}, {Fissel}, \& {Poidevin}}]{Jow+18}
{Jow}, D.~L., {Hill}, R., {Scott}, D., {et~al.} 2018, \mnras, 474, 1018

\bibitem[{{Keown} {et~al.}(2017){Keown}, {Di Francesco}, {Kirk}, {Friesen},
  {Pineda}, {Rosolowsky}, {Ginsburg}, {Offner}, {Caselli}, {Alves},
  {Chac{\'o}n-Tanarro}, {Punanova}, {Redaelli}, {Seo}, {Matzner}, {Chun-Yuan
  Chen}, {Goodman}, {Chen}, {Shirley}, {Singh}, {Arce}, {Martin}, \&
  {Myers}}]{2017ApJ...850....3K}
{Keown}, J., {Di Francesco}, J., {Kirk}, H., {et~al.} 2017, \apj, 850, 3

\bibitem[{{Kerr} {et~al.}(2019){Kerr}, {Kirk}, {Di Francesco}, {Keown}, {Chen},
  {Rosolowsky}, {Offner}, {Friesen}, {Pineda}, {Shirley}, {Redaelli},
  {Caselli}, {Punanova}, {Seo}, {Alves}, {Chac{\'o}n-Tanarro}, \&
  {Chen}}]{kerr_2019}
{Kerr}, R., {Kirk}, H., {Di Francesco}, J., {et~al.} 2019, \apj, 874, 147

\bibitem[{{King} {et~al.}(2019){King}, {Chen}, {Fissel}, \&
  {Li}}]{2019MNRAS.490.2760K}
{King}, P.~K., {Chen}, C.-Y., {Fissel}, L.~M., \& {Li}, Z.-Y. 2019, \mnras,
  490, 2760

\bibitem[{{King} {et~al.}(2018){King}, {Fissel}, {Chen}, \&
  {Li}}]{2018MNRAS.474.5122K}
{King}, P.~K., {Fissel}, L.~M., {Chen}, C.-Y., \& {Li}, Z.-Y. 2018, \mnras,
  474, 5122

\bibitem[{{Kirk} {et~al.}(2017){Kirk}, {Friesen}, {Pineda}, {Rosolowsky},
  {Offner}, {Matzner}, {Myers}, {Di Francesco}, {Caselli}, {Alves},
  {Chac{\'o}n-Tanarro}, {Chen}, {Chun-Yuan Chen}, {Keown}, {Punanova}, {Seo},
  {Shirley}, {Ginsburg}, {Hall}, {Singh}, {Arce}, {Goodman}, {Martin}, \&
  {Redaelli}}]{Kirk+17}
{Kirk}, H., {Friesen}, R.~K., {Pineda}, J.~E., {et~al.} 2017, \apj, 846, 144

\bibitem[{{Kirk} {et~al.}(2013){Kirk}, {Ward-Thompson}, {Palmeirim},
  {Andr{\'e}}, {Griffin}, {Hargrave}, {K{\"o}nyves}, {Bernard}, {Nutter},
  {Sibthorpe}, {Di Francesco}, {Abergel}, {Arzoumanian}, {Benedettini},
  {Bontemps}, {Elia}, {Hennemann}, {Hill}, {Men'shchikov}, {Motte},
  {Nguyen-Luong}, {Peretto}, {Pezzuto}, {Rygl}, {Sadavoy}, {Schisano},
  {Schneider}, {Testi}, \& {White}}]{Kirk+13}
{Kirk}, J.~M., {Ward-Thompson}, D., {Palmeirim}, P., {et~al.} 2013, \mnras,
  432, 1424

\bibitem[{{K{\"o}nyves} {et~al.}(2010){K{\"o}nyves}, {Andr{\'e}},
  {Men'shchikov}, {Schneider}, {Arzoumanian}, {Bontemps}, {Attard}, {Motte},
  {Didelon}, {Maury}, {Abergel}, {Ali}, {Baluteau}, {Bernard}, {Cambr{\'e}sy},
  {Cox}, {di Francesco}, {di Giorgio}, {Griffin}, {Hargrave}, {Huang}, {Kirk},
  {Li}, {Martin}, {Minier}, {Molinari}, {Olofsson}, {Pezzuto}, {Russeil},
  {Roussel}, {Saraceno}, {Sauvage}, {Sibthorpe}, {Spinoglio}, {Testi},
  {Ward-Thompson}, {White}, {Wilson}, {Woodcraft}, \& {Zavagno}}]{Konyves+10}
{K{\"o}nyves}, V., {Andr{\'e}}, P., {Men'shchikov}, A., {et~al.} 2010, \aap,
  518, L106

\bibitem[{{Koyama} \& {Inutsuka}(2000)}]{2000ApJ...532..980K}
{Koyama}, H., \& {Inutsuka}, S.-I. 2000, \apj, 532, 980

\bibitem[{{Kudoh} \& {Basu}(2011)}]{KudohBasu2011}
{Kudoh}, T., \& {Basu}, S. 2011, \apj, 728, 123

\bibitem[{{Lazarian}(2007)}]{Lazarian07}
{Lazarian}, A. 2007, \jqsrt, 106, 225

\bibitem[{{Lazarian} \& {Hoang}(2007)}]{LH07}
{Lazarian}, A., \& {Hoang}, T. 2007, \mnras, 378, 910

\bibitem[{{Lee} \& {Draine}(1985)}]{1985ApJ...290..211L}
{Lee}, H.~M., \& {Draine}, B.~T. 1985, \apj, 290, 211

\bibitem[{{Lee} {et~al.}(2017){Lee}, {Hull}, \& {Offner}}]{LeeHull+17}
{Lee}, J. W.~Y., {Hull}, C. L.~H., \& {Offner}, S. S.~R. 2017, \apj, 834, 201

\bibitem[{{Li} {et~al.}(2014){Li}, {Banerjee}, {Pudritz}, {J{\o}rgensen},
  {Shang}, {Krasnopolsky}, \& {Maury}}]{Li2014PPVI}
{Li}, Z.~Y., {Banerjee}, R., {Pudritz}, R.~E., {et~al.} 2014, in Protostars and
  Planets VI, ed. H.~{Beuther}, R.~S. {Klessen}, C.~P. {Dullemond}, \&
  T.~{Henning}, 173

\bibitem[{{Li} \& {Nakamura}(2004)}]{LiNakamura04}
{Li}, Z.-Y., \& {Nakamura}, F. 2004, \apjl, 609, L83

\bibitem[{{Matthews} {et~al.}(2009){Matthews}, {McPhee}, {Fissel}, \&
  {Curran}}]{Matthews09}
{Matthews}, B.~C., {McPhee}, C.~A., {Fissel}, L.~M., \& {Curran}, R.~L. 2009,
  \apjs, 182, 143

\bibitem[{{Matthews} {et~al.}(2014){Matthews}, {Ade}, {Angil{\`e}}, {Benton},
  {Chapin}, {Chapman}, {Devlin}, {Fissel}, {Fukui}, {Gandilo}, {Gundersen},
  {Hargrave}, {Klein}, {Korotkov}, {Moncelsi}, {Mroczkowski}, {Netterfield},
  {Novak}, {Nutter}, {Olmi}, {Pascale}, {Poidevin}, {Savini}, {Scott},
  {Shariff}, {Soler}, {Tachihara}, {Thomas}, {Truch}, {Tucker}, {Tucker}, \&
  {Ward-Thompson}}]{2014ApJ...784..116M}
{Matthews}, T.~G., {Ade}, P. A.~R., {Angil{\`e}}, F.~E., {et~al.} 2014, \apj,
  784, 116

\bibitem[{{McKee} {et~al.}(2010){McKee}, {Li}, \& {Klein}}]{McKee+10}
{McKee}, C.~F., {Li}, P.~S., \& {Klein}, R.~I. 2010, \apj, 720, 1612

\bibitem[{{McKee} \& {Ostriker}(2007)}]{MO2007}
{McKee}, C.~F., \& {Ostriker}, E.~C. 2007, \araa, 45, 565

\bibitem[{{Mestel}(1985)}]{1985prpl.conf..320M}
{Mestel}, L. 1985, in Protostars and Planets II, ed. D.~C. {Black} \& M.~S.
  {Matthews}, 320--339

\bibitem[{{Mestel} \& {Spitzer}(1956)}]{1956MNRAS.116..503M}
{Mestel}, L., \& {Spitzer}, L., J. 1956, \mnras, 116, 503

\bibitem[{{Monsch} {et~al.}(2018){Monsch}, {Pineda}, {Liu}, {Zucker}, {How-Huan
  Chen}, {Pattle}, {Offner}, {Di Francesco}, {Ginsburg}, {Ercolano}, {Arce},
  {Friesen}, {Kirk}, {Caselli}, \& {Goodman}}]{Monsch+18}
{Monsch}, K., {Pineda}, J.~E., {Liu}, H.~B., {et~al.} 2018, \apj, 861, 77

\bibitem[{{Novak} {et~al.}(1997){Novak}, {Dotson}, {Dowell}, {Goldsmith},
  {Hildebrand}, {Platt}, \& {Schleuning}}]{Novak1997}
{Novak}, G., {Dotson}, J.~L., {Dowell}, C.~D., {et~al.} 1997, \apj, 487, 320

\bibitem[{{Onishi} {et~al.}(1996){Onishi}, {Mizuno}, {Kawamura}, {Ogawa}, \&
  {Fukui}}]{Onishi96}
{Onishi}, T., {Mizuno}, A., {Kawamura}, A., {Ogawa}, H., \& {Fukui}, Y. 1996,
  \apj, 465, 815

\bibitem[{{Ostriker} {et~al.}(1999){Ostriker}, {Gammie}, \&
  {Stone}}]{Ostriker+99}
{Ostriker}, E.~C., {Gammie}, C.~F., \& {Stone}, J.~M. 1999, \apj, 513, 259

\bibitem[{{Padovani} {et~al.}(2012){Padovani}, {Brinch}, {Girart},
  {J{\o}rgensen}, {Frau}, {Hennebelle}, {Kuiper}, {Vlemmings}, {Bertoldi},
  {Hogerheijde}, {Juhasz}, \& {Schaaf}}]{2012A&A...543A..16P}
{Padovani}, M., {Brinch}, C., {Girart}, J.~M., {et~al.} 2012, \aap, 543, A16

\bibitem[{{Pascale} {et~al.}(2012){Pascale}, {Ade}, {Angil{\`e}}, {Benton},
  {Devlin}, {Dober}, {Fissel}, {Fukui}, {Gandilo}, {Gundersen}, {Hargrave},
  {Klein}, {Korotkov}, {Matthews}, {Moncelsi}, {Mroczkowski}, {Netterfield},
  {Novak}, {Nutter}, {Olmi}, {Poidevin}, {Savini}, {Scott}, {Shariff}, {Soler},
  {Thomas}, {Truch}, {Tucker}, {Tucker}, \&
  {Ward-Thompson}}]{2012SPIE.8444E..15P}
{Pascale}, E., {Ade}, P. A.~R., {Angil{\`e}}, F.~E., {et~al.} 2012, Society of
  Photo-Optical Instrumentation Engineers (SPIE) Conference Series, Vol. 8444,
  {The balloon-borne large-aperture submillimeter telescope for
  polarimetry-BLASTPol: performance and results from the 2010 Antarctic
  flight}, 844415

\bibitem[{{Pineda} {et~al.}(2011){Pineda}, {Goodman}, {Arce}, {Caselli},
  {Longmore}, \& {Corder}}]{Pineda+11}
{Pineda}, J.~E., {Goodman}, A.~A., {Arce}, H.~G., {et~al.} 2011, \apjl, 739, L2

\bibitem[{{Pineda} {et~al.}(2015){Pineda}, {Offner}, {Parker}, {Arce},
  {Goodman}, {Caselli}, {Fuller}, {Bourke}, \& {Corder}}]{Pineda+15}
{Pineda}, J.~E., {Offner}, S. S.~R., {Parker}, R.~J., {et~al.} 2015, \nat, 518,
  213

\bibitem[{{Planck Collaboration Int. XIX}(2015)}]{planckXIX}
{Planck Collaboration Int. XIX}. 2015, \aap, 576, A104

\bibitem[{{Planck Collaboration Int. XX}.(2015)}]{PlanckXX}
{Planck Collaboration Int. XX}. 2015, \aap, 576, A105

\bibitem[{{Planck Collaboration Int. XXXIII}.(2016)}]{pXXXIII}
{Planck Collaboration Int. XXXIII}. 2016, \aap, 586, A136

\bibitem[{{Planck Collaboration Int. XXXV}(2016)}]{pxxxv16}
{Planck Collaboration Int. XXXV}. 2016, \aap, 586, A138

\bibitem[{{Poidevin} {et~al.}(2014){Poidevin}, {Ade}, {Angile}, {Benton},
  {Chapin}, {Devlin}, {Fissel}, {Fukui}, {Gand ilo}, {Gundersen}, {Hargrave},
  {Klein}, {Korotkov}, {Matthews}, {Moncelsi}, {Mroczkowski}, {Netterfield},
  {Novak}, {Nutter}, {Olmi}, {Pascale}, {Savini}, {Scott}, {Shariff}, {Diego
  Soler}, {Tachihara}, {Thomas}, {Truch}, {Tucker}, {Tucker}, \&
  {Ward-Thompson}}]{2014ApJ...791...43P}
{Poidevin}, F., {Ade}, P. A.~R., {Angile}, F.~E., {et~al.} 2014, \apj, 791, 43

\bibitem[{{Punanova} {et~al.}(2018){Punanova}, {Caselli}, {Pineda}, {Pon},
  {Tafalla}, {Hacar}, \& {Bizzocchi}}]{Punanova+18}
{Punanova}, A., {Caselli}, P., {Pineda}, J.~E., {et~al.} 2018, \aap, 617, A27

\bibitem[{{Rathborne} {et~al.}(2009){Rathborne}, {Lada}, {Muench}, {Alves},
  {Kainulainen}, \& {Lombardi}}]{2009ApJ...699..742R}
{Rathborne}, J.~M., {Lada}, C.~J., {Muench}, A.~A., {et~al.} 2009, \apj, 699,
  742

\bibitem[{{Rosolowsky} \& {Leroy}(2006)}]{2006PASP..118..590R}
{Rosolowsky}, E., \& {Leroy}, A. 2006, \pasp, 118, 590

\bibitem[{{Rygl} {et~al.}(2013){Rygl}, {Benedettini}, {Schisano}, {Elia},
  {Molinari}, {Pezzuto}, {Andr{\'e}}, {Bernard}, {White}, {Polychroni},
  {Bontemps}, {Cox}, {Di Francesco}, {Facchini}, {Fallscheer}, {di Giorgio},
  {Hennemann}, {Hill}, {K{\"o}nyves}, {Minier}, {Motte}, {Nguyen-Luong},
  {Peretto}, {Pestalozzi}, {Sadavoy}, {Schneider}, {Spinoglio}, {Testi}, \&
  {Ward-Thompson}}]{2013A&A...549L...1R}
{Rygl}, K.~L.~J., {Benedettini}, M., {Schisano}, E., {et~al.} 2013, \aap, 549,
  L1

\bibitem[{{Seo} {et~al.}(2015){Seo}, {Shirley}, {Goldsmith}, {Ward-Thompson},
  {Kirk}, {Schmalzl}, {Lee}, {Friesen}, {Langston}, {Masters}, \&
  {Garwood}}]{seo_2015}
{Seo}, Y.~M., {Shirley}, Y.~L., {Goldsmith}, P., {et~al.} 2015, \apj, 805, 185

\bibitem[{{Shirley}(2015)}]{2015PASP..127..299S}
{Shirley}, Y.~L. 2015, \pasp, 127, 299

\bibitem[{{Shu} {et~al.}(1987){Shu}, {Adams}, \& {Lizano}}]{Shu1987}
{Shu}, F.~H., {Adams}, F.~C., \& {Lizano}, S. 1987, \araa, 25, 23

\bibitem[{{Soler}(2019)}]{Soler19}
{Soler}, J.~D. 2019, \aap, 629, A96

\bibitem[{{Soler} {et~al.}(2013){Soler}, {Hennebelle}, {Martin},
  {Miville-Desch{\^e}nes}, {Netterfield}, \& {Fissel}}]{Soler+13}
{Soler}, J.~D., {Hennebelle}, P., {Martin}, P.~G., {et~al.} 2013, \apj, 774,
  128

\bibitem[{{Soler} {et~al.}(2019){Soler}, {Beuther}, {Rugel}, {Wang}, {Clark},
  {Glover}, {Goldsmith}, {Heyer}, {Anderson}, {Goodman}, {Henning},
  {Kainulainen}, {Klessen}, {Longmore}, {McClure-Griffiths}, {Menten},
  {Mottram}, {Ott}, {Ragan}, {Smith}, {Urquhart}, {Bigiel}, {Hennebelle},
  {Roy}, \& {Schilke}}]{Soler+19}
{Soler}, J.~D., {Beuther}, H., {Rugel}, M., {et~al.} 2019, \aap, 622, A166

\bibitem[{{Stephens} {et~al.}(2017){Stephens}, {Dunham}, {Myers}, {Pokhrel},
  {Sadavoy}, {Vorobyov}, {Tobin}, {Pineda}, {Offner}, {Lee}, {Kristensen},
  {J{\o}rgensen}, {Goodman}, {Bourke}, {Arce}, \& {Plunkett}}]{Stephens+17}
{Stephens}, I.~W., {Dunham}, M.~M., {Myers}, P.~C., {et~al.} 2017, \apj, 846,
  16

\bibitem[{{Storm} {et~al.}(2014){Storm}, {Mundy}, {Fern{\'a}ndez-L{\'o}pez},
  {Lee}, {Looney}, {Teuben}, {Rosolowsky}, {Arce}, {Ostriker}, {Segura-Cox},
  {Pound}, {Salter}, {Volgenau}, {Shirley}, {Chen}, {Gong}, {Plunkett},
  {Tobin}, {Kwon}, {Isella}, {Kauffmann}, {Tassis}, {Crutcher}, {Gammie}, \&
  {Testi}}]{CLASSy2014}
{Storm}, S., {Mundy}, L.~G., {Fern{\'a}ndez-L{\'o}pez}, M., {et~al.} 2014,
  \apj, 794, 165

\bibitem[{{Sullivan} {et~al.}(2019){Sullivan}, {Fissel}, {King}, {Chen}, {Li},
  \& {Soler}}]{Colin+19}
{Sullivan}, C., {Fissel}, L.~M., {King}, P.~K., {et~al.} 2019, submitted to
  \mnras

\bibitem[{{Van Loo} {et~al.}(2014){Van Loo}, {Keto}, \&
  {Zhang}}]{2014ApJ...789...37V}
{Van Loo}, S., {Keto}, E., \& {Zhang}, Q. 2014, \apj, 789, 37

\bibitem[{{V{\'a}zquez-Semadeni} {et~al.}(2006){V{\'a}zquez-Semadeni}, {Ryu},
  {Passot}, {Gonz{\'a}lez}, \& {Gazol}}]{2006ApJ...643..245V}
{V{\'a}zquez-Semadeni}, E., {Ryu}, D., {Passot}, T., {Gonz{\'a}lez}, R.~F., \&
  {Gazol}, A. 2006, \apj, 643, 245

\bibitem[{{Wilking}(1992)}]{1992lmsf.book..159W}
{Wilking}, B.~A. 1992, {Star Formation in the Ophiuchus Molecular Cloud
  Complex}, ed. B.~{Reipurth}, 159

\bibitem[{{Wilking} {et~al.}(2008){Wilking}, {Gagn{\'e}}, \&
  {Allen}}]{2008hsf2.book..351W}
{Wilking}, B.~A., {Gagn{\'e}}, M., \& {Allen}, L.~E. 2008, {Star Formation in
  the {\ensuremath{\rho}} Ophiuchi Molecular Cloud}, ed. B.~{Reipurth}, Vol.~5,
  351

\bibitem[{{Yang} {et~al.}(2016){Yang}, {Li}, {Looney}, \&
  {Stephens}}]{Haifeng16}
{Yang}, H., {Li}, Z.-Y., {Looney}, L., \& {Stephens}, I. 2016, \mnras, 456,
  2794

\end{thebibliography}


\appendix

\section{Cores Identified in GAS Data and Their Relative Alignments}
\label{sec::GASmaps}

\begin{figure*}
    \centering
    \includegraphics[width=\textwidth]{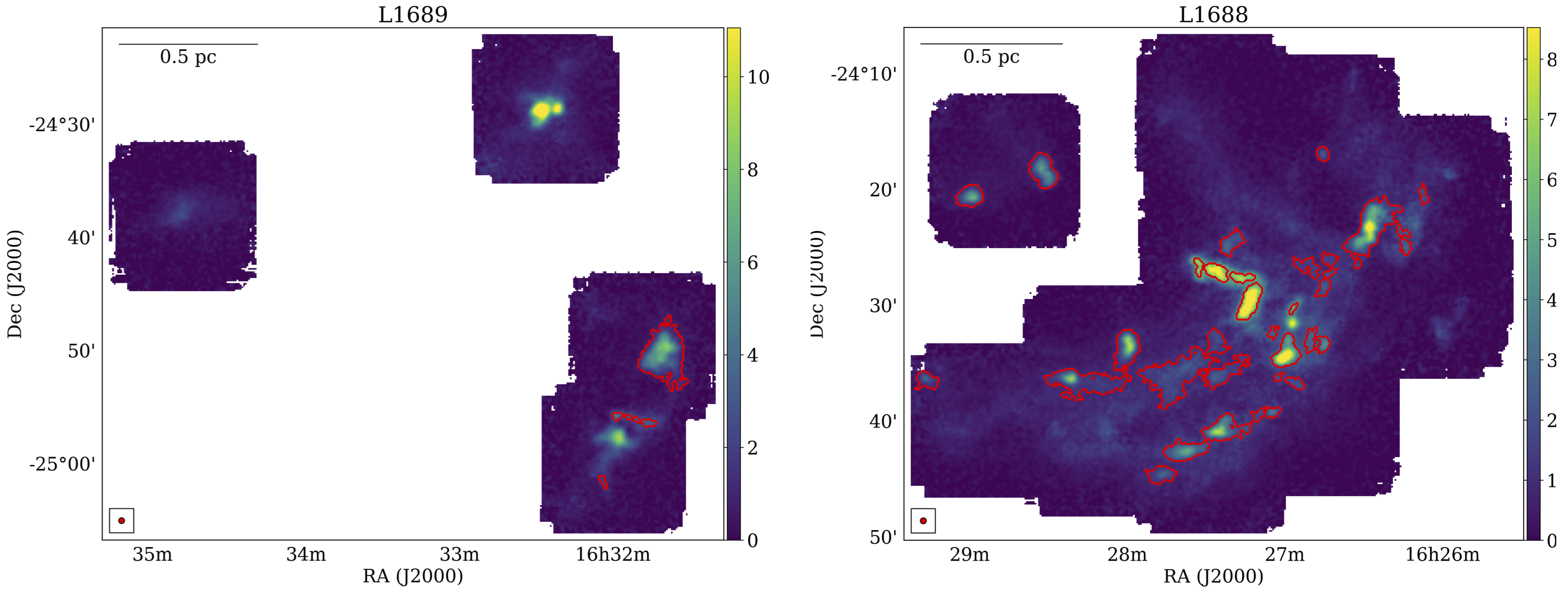}
    \caption{Cores ({\it red contours}) identified in each individual region in the Ophiuchus cloud using \amm{} integrated emission ({\it colormaps}, in K\,km\,s$^{-1}$) observed by GAS. Note that cores that are more rounded (with minor/major ratio $> 0.8$ are removed from the analysis and not shown here. Dendrogram leaves without obvious central peak emission are also excluded. }
    \label{fig:mapOph}
\end{figure*}

\begin{figure*}
    \centering
    \includegraphics[width=\textwidth]{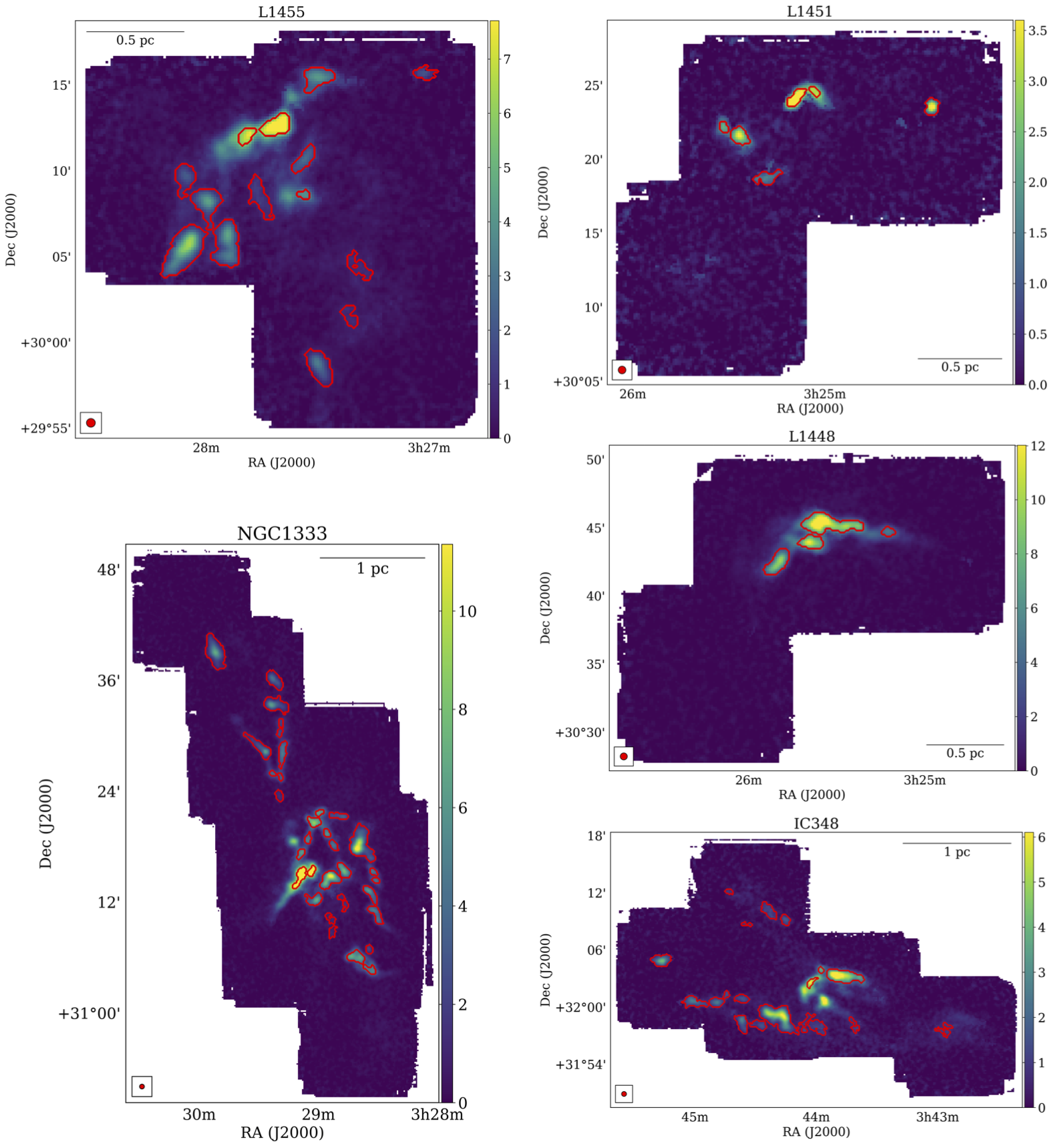}
    \caption{Same as Figure~\ref{fig:mapOph}, but for the Perseus cloud.}
    \label{fig:mapPer}
\end{figure*}

\begin{figure*}
    \centering
    \includegraphics[width=\textwidth]{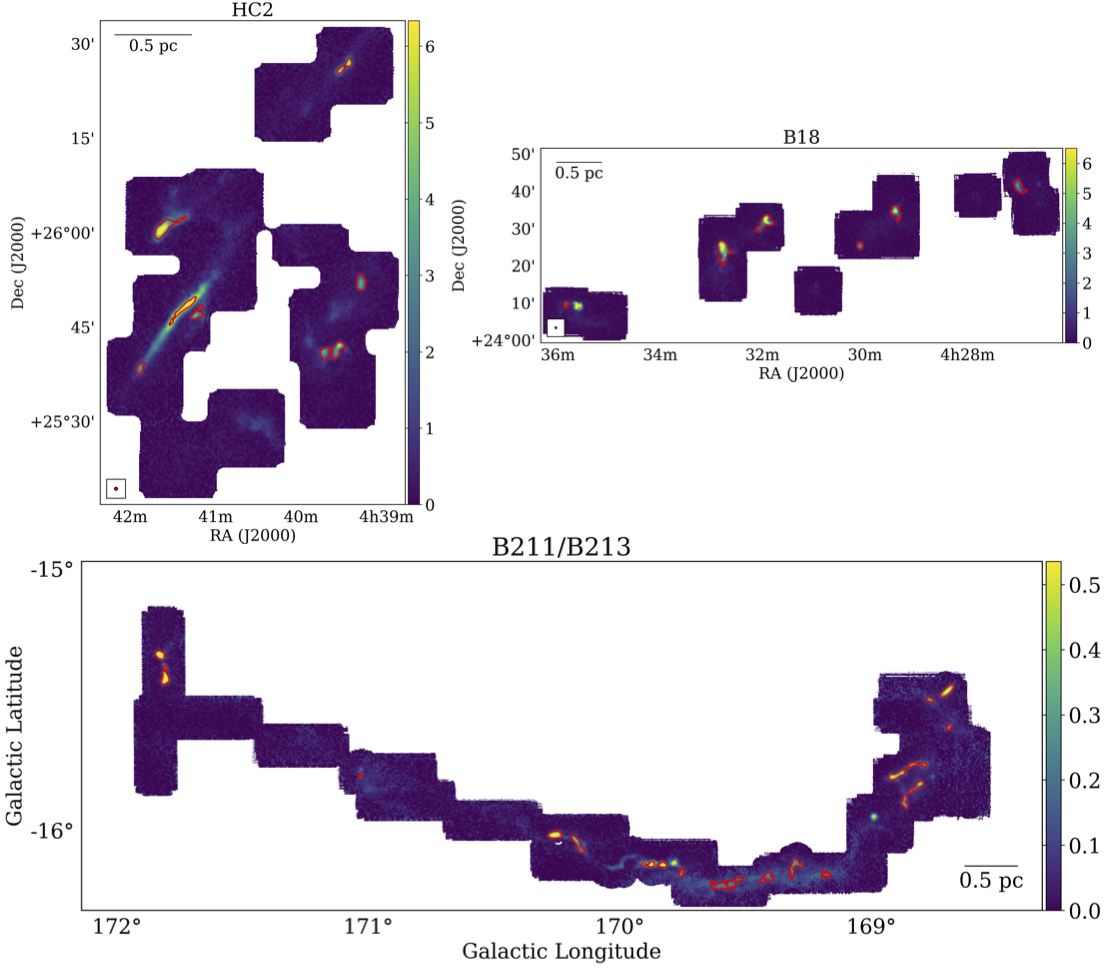}
    \caption{Same as Figure~\ref{fig:mapOph}, but for the Taurus cloud.}
    \label{fig:mapTau}
\end{figure*}

\begin{figure*}
    \centering
    \includegraphics[width=0.3\textwidth]{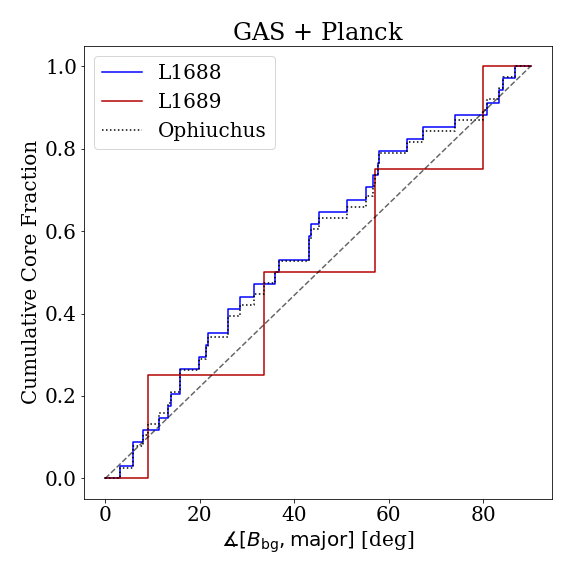}
    \includegraphics[width=0.3\textwidth]{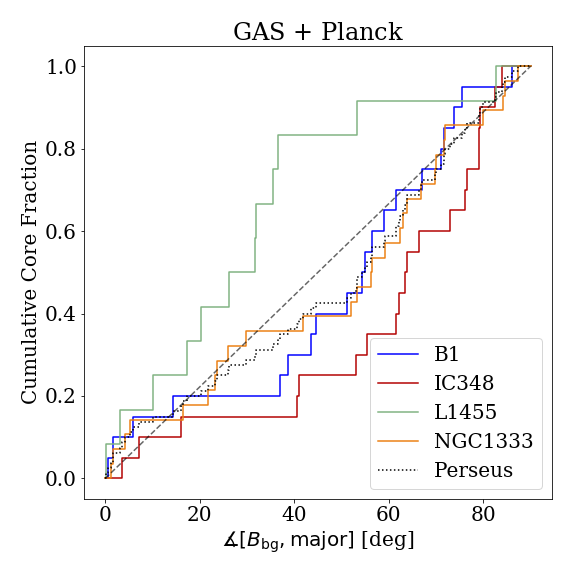}
    \includegraphics[width=0.3\textwidth]{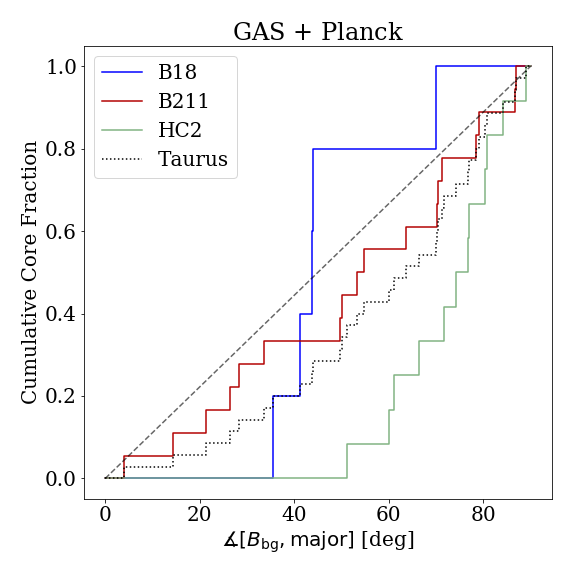}
    \caption{The cumulative distributions of the core--background magnetic field alignment for the sub-regions within the three clouds: Ophiuchus ({\it left}), Perseus ({\it middle}), and Taurus ({\it right}).}
    \label{fig:regStep}
\end{figure*}

Figures~\ref{fig:mapOph}$-$\ref{fig:mapTau} show all sub-regions observed by GAS overlapped with the core boundaries identified by \texttt{astrodendro}. The relative alignment between cores and background magnetic field (i.e.,~the angle between core major axis and background magnetic field direction) in each sub-region is plotted in Figure~\ref{fig:regStep}.

\section{Relative Angles in 2D and 3D}
\label{sec::apx}

\begin{figure}
    \centering
    \includegraphics[width=\columnwidth]{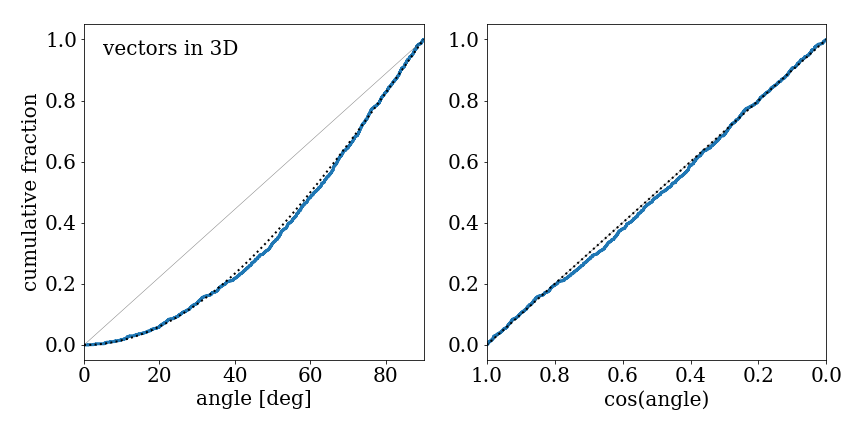}
    \includegraphics[width=\columnwidth]{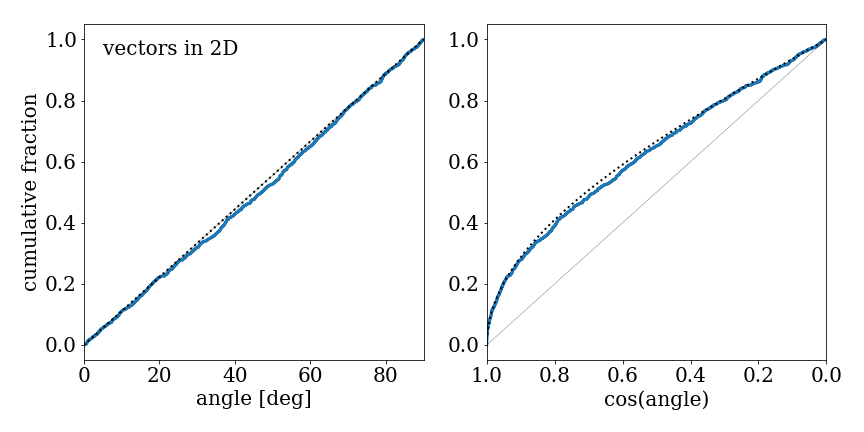}
    \caption{Comparing the relative angles in 3D ({\it top}) and 2D ({\it bottom}), for both analytic solutions of Equations~(\ref{eq::CP3D}) and (\ref{eq::CP2D}) ({\it black dotted lines}) and between two sets of random vectors ({\it solid lines}), showing the angle in degree ({\it left}) and the cosine values ({\it right}). The grey diagonal lines represent random distributions. }
    \label{fig:randist}
\end{figure}

Here we briefly describe the difference between measuring relative angle between two random vectors in 3D and 2D spaces. 

Without loss of generality, we assume both vectors are unit vectors and one is pointing straight up ($+z$ for 3D case and $+y$ for 2D case). The probability for the second vector to have an angle $\theta$ with respect to the first one is therefore proportional to the surface area of the stripe of constant latitude $\theta$ on the 3D sphere ($2\pi R\sin\theta\cdot Rd\theta$), or the arc length at $\theta$ of the unit circle ($R d\theta$). For unit vectors, $R=1$, and since we only consider angles within $[0,90]$ degree, the cumulative probability of the relative angle to be smaller than $\theta$ is therefore
\begin{align}
    \int_0^\theta \frac{2\pi\sin\theta}{2\pi} d\theta &= 1-\cos\theta\ \ \ \mathrm{(3D\ case)},\label{eq::CP3D}\\
    \int_0^\theta \frac{2}{\pi}d\theta &= 2\frac{\theta}{\pi}\ \ \ \mathrm{(2D\ case)}.\label{eq::CP2D}
\end{align}
Note that the probability is normalized by the total surface area of the sphere (in 3D) or the circumference of the whole circle (in 2D).
This explains why a set of 3D random vectors will show a linear distribution in $\cos\alpha$ instead of $\alpha$, where $\alpha$ is the relative angle between any two vectors from the set.
Equations~(\ref{eq::CP3D}) and (\ref{eq::CP2D}) are illustrated in Figure~\ref{fig:randist} as black dotted lines, which also shows that two vectors in 3D are more likely to be perpendicular than parallel if measured in degrees.

 We further present a simple numerical test to demonstrate the fact.
We use the \texttt{numpy.random.rand} function in \texttt{python} to generate six sets of 1,000 random numbers and subtract by their mean values so that the new mean values are equal to 0. These are the $x$, $y$, and $z$ components of two sets of 1,000 randomly pointed vectors $u$ and $v$.
We then calculate the relative angle between the individual vectors from each of the sets, $\cos\alpha = |u_xv_x + u_yv_y + u_zv_z|/|u||v|$. Similar to our analysis presented in this paper, we limited the relative angle $\alpha$ to be within $[0,90]$ degrees. 

The resulting cumulative distribution (in step function style) of $\alpha$ is shown in Figure~\ref{fig:randist} (top panels), measured in both $\alpha$ itself (left) and in $\cos\alpha$ (right). Note that as in Figures~\ref{fig:Bbg}, \ref{fig:Bmajor}, and \ref{fig:Bgmajor}, the $x$ axis of the cosine plot has been flipped so that the shape of the cumulative distribution can be directly compared to the plot in angles. 
It is clear that even though these 1,000 measurements of relative angle between two randomly-pointed vectors are supposed to be random, their distribution in angle does not reflect that, and shows a preference for perpendicular alignment instead. On the other hand, the cumulative distribution in cosine value of such relative angle perfectly follows the expected shape of random distribution (a straight diagonal line in cumulative plot).

For comparison, we repeated the same analysis for vectors in 2D. We again generated four sets of 1,000 random values centered at 0 and used them to create two sets of randomly pointed vectors in 2D. The cumulative distribution of the relative angles between vectors from these two sets is shown in Figure~\ref{fig:randist} (bottom panel). For these 2D cases, the measurements in angle follow the expected random distribution very well, while the cosine values suggest a moderate preference toward small angles. Combining with Equations~(\ref{eq::CP3D}) and (\ref{eq::CP2D}), Figure~\ref{fig:randist} therefore provides justifications for our choices of angle measurements presented in this paper, and can be used as a guideline when interpreting our results.


\bsp	
\label{lastpage}
\end{document}